\documentclass[vecphys]{svmult}
\usepackage{makeidx}         
\usepackage{graphicx}        
\usepackage{multicol}        
\usepackage[bottom]{footmisc}
\makeindex             

\begin{document}

\title*{Dynamics of small bodies in planetary systems}
\author{Mark C. Wyatt}
\institute{Institute of Astronomy, University of Cambridge,
Cambridge CB3 0HA, UK
\texttt{wyatt@ast.cam.ac.uk}
}
\maketitle

The number of stars that are known to have debris disks is greater than
that of stars known to harbour planets.
These disks are detected because dust is created in
the destruction of planetesimals in the disks much in the same way
that dust is produced in the asteroid belt and Kuiper belt in the solar system.
For the nearest stars the structure of their debris disks can be directly
imaged, showing a wide variety of both axisymmetric and asymmetric
structures.
A successful interpretation of these images requires a knowledge of the
dynamics of small bodies in planetary systems, since this allows the
observed dust distribution to be deconvolved to provide information on
the distribution of larger objects, such as planetesimals and planets.
This chapter reviews the structures seen in debris disks, and describes
a disk dynamical theory which can be used to interpret those observations.
In this way much of the observed structures, both axisymmetric and asymmetric,
can be explained by a model in which the dust is produced in a planetesimal belt
which is perturbed by a nearby, as yet unseen, planet.
While the planet predictions still require confirmation, it is clear that
debris disks have the potential to provide unique information about the
structure of extrasolar planetary systems, since they can tell us about
planets analogous to Neptune and even the Earth.
Significant failings of the model at present are its inability to predict
the quantity of small grains in a system, and to explain the origin of the
transient dust seen in some systems.
Given the complexity of planetary system dynamics and how that is
readily reflected in the structure of a debris disk, it seems inevitable
that the study of debris disks will play a vital role in our
understanding of extrasolar planetary systems.

\section{Introduction}
\label{s:intro}
Planetary systems are not just made up of \textit{planets}, but
are also composed of numerous small bodies ranging from
asteroids and comets as large as 1000 km down to sub-$\mu$m-sized dust
grains.
In the solar system the asteroids and comets are confined to relatively
narrow rings known as the asteroid belt and the Kuiper belt (see chapters
by Nakamura and Jewitt).
These belts are the source of the majority of the smaller objects
seen in the solar system, since such objects are inevitably created
in collisions between objects within the belts (see chapter by Michel).
Sublimation of comets as they are heated on approach to the Sun
is another source of dust in the solar system.

It is known that extrasolar systems also host belts of planetesimals
(a generic name for comets and asteroids) that are similar to our
own asteroid belt and Kuiper belt.
These were first discovered using far-IR observations of nearby
stars, which showed excess emission above that expected to come from
the stellar photosphere \cite{auma84}.
This emission comes from dust that is heated by
the star and which re-radiates that energy in the thermal infrared,
at temperatures between 40-200 K, depending on the distance of the
dust from the star.
The lifetime of the dust is inferred to be short compared with the
age of the star, and so it is concluded that the dust cannot be a
remnant of the proto-planetary disk that formed with the star
(see chapter by Takeuchi), rather it must originate in planetesimal
belts much in the same way that dust is created
in the solar system \cite{bp93}.

Over 300 main sequence stars are now known with this type of excess
emission \cite{mb98,su06,bryd06}, and such objects are either 
known as Vega-like (after the
first star discovered to have this excess), or as debris disks.
Statistical studies have shown that $\sim 15$\% of normal main sequence
stars have debris disks, although it should be stressed that the
disks which can be detected with current technology have greater quantities
of dust than is currently present in the solar system by a factor of at least
10 \cite{gwhd04}.
Nevertheless this indicates that debris disks are common, more common
in fact that extrasolar planets which are found around $\sim 6$\% of
stars \cite{fv05}.
Studying these disks provides a
unique insight into the structure of the planetary systems of other stars.
Indeed, the nearest and brightest debris disks can be imaged, and
such studies have provided the first images of nearby planetary systems.
These images reveal the distribution of dust in the systems, which can
in turn be used to infer the distribution of parent planetesimals, and
also the architecture of the underlying planetary system.
However, to do so requires an understanding of both the mechanism by
which dust is produced in planetesimal belts and its consequent
dynamical evolution, as well as of the dynamical interaction between
planets and planetesimals and between planets and dust.

This chapter reviews our knowledge of debris disks from observations
(\S \ref{s:obs}) and describes a simple model for planetesimal
belt evolution which explains what we see (\S \ref{s:mod}), as well as 
how the detailed interaction between planets and planetesimals
imposes structure on that planetesimal belt (\S \ref{s:plpl}), and
how those perturbations translate into structures seen in the dust
distribution (\S \ref{s:pldust}).
Conclusions, including what has been learned about the planetary systems
of nearby stars from studying these disks, are given in \S \ref{s:conc}.

\section{Observed debris disk structures}
\label{s:obs}
The debris disks with well resolved structure are summarised in
Table \ref{tab:disks}. \footnote{Resolved disks have also been reported
for the following stars: HD92945, HD61005, HD10647, HD202917, and
HD207129 (see http://astro.berkeley.edu/$\sim$kalas/lyot2007/agenda.html),
and HD15745 (Kalas et al., ApJ, submitted).
I have excluded these images from the discussion, since they have yet to
appear in the literature at the time of writing.}
There are two types of debris disk structure:
axisymmetric structure (i.e., dust or planetesimal surface density
as a function of distance from the star), and
asymmetric structure (i.e., how that surface density varies as a
function of azimuth).
I will deal with each of these in turn.

\begin{table}
\centering
\caption{Summary of observed properties of debris disks the structure
of which has been significantly resolved at one wavelength or more.
Asymmetries are identified as: W=Warp, C=Clump, S=Spiral, B=Brightness asymmetry,
O=Offset, H=Hot dust component, N=No discernible asymmetry.}
\label{tab:disks}
\begin{tabular}{ccc|ccccc}
\hline\noalign{\smallskip}
Name & Sp Type & Age, Myr & $r$, AU & $i$, $^\circ$ & $f=L_{\rm{ir}}/L_\star$
& Asymm & Ref \\
\noalign{\smallskip}\hline\noalign{\smallskip}
HD141569           & B9.5e & 5     & $34-1200$  & 35       & $84 \times 10^{-4}$   & S     & \cite{ftpk00,clam03} \\ 
HR4796             & A0V   & 8     & $60-80$    & 17       & $50 \times 10^{-4}$   & B     & \cite{ssbk99,tele00} \\ 
$\beta$ Pictoris   & A5V   & 12    & $10-1835$  & $\sim 3$ & $26 \times 10^{-4}$   & WC    & \cite{hgzw98,hllc00,tele05} \\ 
HD15115            & F2V   & 12    & $31-554$   & $\sim 0$ & $5 \times 10^{-4}$    & B     & \cite{kfg07} \\ 
HD181327           & K2V   & 12    & $68-104$   & 58       & $25 \times 10^{-4}$   & N     & \cite{schn06} \\ 
AU Mic             & M1Ve  & 12    & $12-210$   & $\sim 0$ & $6 \times 10^{-4}$    & WC    & \cite{liu04} \\ 
HD32297            & A0V   & $<30$ & $40-1680$  & 10       & $27 \times 10^{-4}$   & B     & \cite{kala05} \\ 
HD107146           & G2V   & 100   & $80-185$   & 65       & $12 \times 10^{-4}$   & N     & \cite{ardi04} \\ 
HD92945            & K1V   & 100   & $45-175$   & 29       & $8 \times 10^{-4}$    & N     & \cite{goli07} \\ 
Fomalhaut          & A3V   & 200   & $133-158$  & 24       & $0.8 \times 10^{-4}$  & (CB)O & \cite{hgdw03,stap04,kgc05} \\ 
HD139664           & F5V   & 300   & $83-109$   & $<5$     & $0.9 \times 10^{-4}$  & N     & \cite{kgcf06} \\  
Vega               & A0V   & 350   & $90-800$   & $\sim 90$& $0.2 \times 10^{-4}$  & C     & \cite{hgzw98,kso01,su05,mdvg06} \\
$\epsilon$ Eridani & K2V   & 850   & $40-105$   & 65       & $0.8 \times 10^{-4}$  & CO    & \cite{ghwd05} \\ 
$\eta$ Corvi       & F2V   & 1000  & $1.5$, 150 & 45       & $5.3 \times 10^{-4}$  & CH    & \cite{wgdc05} \\ 
HD53143            & K1V   & 1000  & $55-110$   & 45       & $2.5 \times 10^{-4}$  & N     & \cite{kgcf06} \\  
$\tau$ Ceti        & G2V   & 10000 & $\sim 55$  & 60-90    & $0.3 \times 10^{-4}$  & N     & \cite{gwhd04} \\ 
\noalign{\smallskip}\hline
\end{tabular}
\end{table}

\subsection{Axisymmetric structure}
\label{ss:axi}
The most basic information about the structure of a debris disk that we
can obtain is the distance of the dust from the star.
This can be deduced without resolving the dust location by looking at 
the Spectral Energy Distribution (SED), since this indicates the temperature of 
the dust, which by thermal balance with the stellar luminosity tells us its
distance from the star.
For black body dust
\begin{equation}
  T_{\rm{bb}} = 278.3 L_\star^{0.25}/\sqrt{r},
  \label{eq:tbb}
\end{equation}
where $L_\star$ is in $L_\odot$ and $r$ is distance from star in AU.
Thus dust location, $r$, can be estimated as long as the level
of dust emission has been measured at two or more wavelengths from
which its temperature can be estimated.

However, such estimates suffer large uncertainties, since the exact
temperature of the dust depends on its size and composition
(see chapter by Li).
Assuming black body emission for the grains can underestimate 
(or overestimate) the distance
of the dust from the star by a factor of 3 or more if the dust is small
\cite{schn06}, since small grains emit inefficiently at long wavelengths and
so attain equilibrium temperatures that are significantly higher than black body
\cite{wdtf99}.
Likewise, an SED which can be fitted by a black body emission spectrum
does not necessarily indicate that all of the dust is at a single
distance from the star, any more than one that requires multiple
temperatures indicates that the disk is broad, since dust at multiple
distances can appear to have one temperature, and dust with a range
of sizes at the same distance from the star have a range of temperatures
\cite{mkw07}. 
This underlines the fact that the interpretation of SEDs is degenerate,
and that in order to determine the radial structure of a disk it needs
to be spatially resolved.
On the other hand, once the radial location of the dust is known the
information in the SED is extremely valuable, since it allows a determination
of the emission properties of the grains, and hence of their size and/or 
composition \cite{wd02,ll03}.

Nevertheless, it seems that the majority of the known debris disks have SEDs
that are dominated by dust at a single temperature, and are seen in images to
be dominated by dust at a distance from the star that is compatible with that
temperature.
More often than not that distance is $>30$ AU from the star, which means
that debris disks are analogous to our own Kuiper belt \cite{whgd03}.
Naturally, the fact that these disks have inner holes similar in size to the
planetary system in our solar system leads to the intriguing possibility that
there is an (as yet) unseen planetary system sweeping these regions free of both
planetesimals and dust.
I will return to the putative planetary system in \S \ref{s:plpl}.
Here I simply note that while these inner holes are usually dust free \cite{wnlb04}, 
a few systems are known with dust within this hole, such as $\eta$ Corvi
\cite{wgdc05} and Vega \cite{absi06}.
The hot dust in these systems is thought to be transiently
regenerated \cite{wsgb07}, and care would need to be taken
when interpreting observations of the hot dust within the framework
described in this chapter (see \S \ref{s:conc}).

Exactly how broad these disks are is a matter for debate.
Optical imaging suggests that there are two types of disk, narrow and broad \cite{kgcf06}.
However, detectability may be an issue in some cases, since a disk's outer
edge is often difficult to detect, as the fraction of intercepted starlight
falls off with radius, much in the same way that it was not known for a long time
whether or not the outer edge of the Kuiper belt is abrupt \cite{tb01}.
Some disks clearly are extended though, such as that of $\beta$ Pictoris,
which is seen to extend out to $>1000$ AU in optical imaging \cite{st84}, but
which is seen as close in as $10-20$ AU in mid-IR imaging \cite{tele05}.
While disks with a dust distribution as broad as that of $\beta$ Pictoris are
rare, the presence of dust at large distances from the star is becoming more
common-place.
It is now known that dust in the archetypal debris disk Vega is not
confined to $\sim 90$ AU as suggested by sub-mm images \cite{hgzw98}.
Rather the dust distribution seen at 24 and 70 $\mu$m extends out to 800 AU
\cite{su05}.
This defies intuition, since if the dust is at a range of distances,
the disk should appear smaller at the shorter wavelengths
(since shorter wavelengths tend to probe hotter dust).
This intuitive behaviour is indeed seen in the disk of Fomalhaut \cite{stap04}.
It is thought that the counter-intuitive behaviour of Vega's disk
arises because the grain size distribution changes with distance:
the dust seen at large distances is small, of order a few $\mu$m,
and so is heated above black body and emits very inefficiently in the sub-mm,
whereas that seen at $\sim 90$ AU is large, mm- to cm-sized, and emits
efficiently in the sub-mm at (relatively low) black body-like temperatures.
A similar change in size distribution with distance is seen in the extended dust 
distributions of $\beta$ Pictoris and AU Mic.
The extension of these disks is not seen in mid- and far-IR images, but in
optical and near-IR images of scattered starlight, and the colours and
polarisation of the scattered light show that the dust at large distances
in these systems is small, sub-$\mu$m in size \cite{gkm07}.

\subsection{Asymmetric structure}
\label{ss:asymm}
\begin{figure}
\centering
  \includegraphics[height=10cm,angle=0]{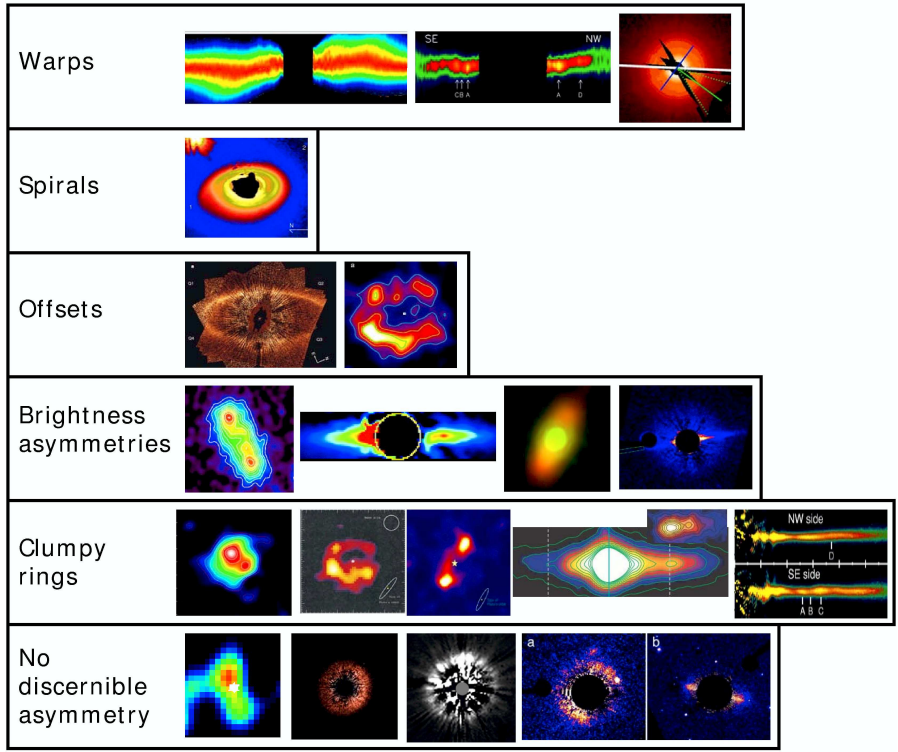}
\caption{Summary of asymmetries seen in the structures of debris disks.
References for the images are from left to right:
  warps ($\beta$ Pictoris \cite{hllc00},
         AU Mic reprinted with permission from AAAS \cite{liu04},
         TW Hydra \cite{rwm05});
  spirals (HD141569 \cite{clam03});
  offsets (Fomalhaut reprinted by permission from Macmillan Publishers
         Ltd: Nature \cite{kgc05} copyright 2005,
         $\epsilon$ Eridani \cite{ghwd05});
  brightness asymmetries (HR4796 \cite{tele00},
         HD32297 \cite{ssh05}, 
         Fomalhaut \cite{stap04},
         HD15115 \cite{kfg07});
  clumpy rings (Vega \cite{hgzw98}, 
         $\epsilon$ Eridani \cite{ghmj98},
         Fomalhaut \cite{hgdw03},
         $\beta$ Pictoris \cite{tele05},
         AU Mic \cite{kris05});
  no discernible asymmetry ($\tau$ Ceti \cite{gwhd04},
         HD107146 \cite{ardi04},
         HD181327 \cite{schn06},
         HD53143 \cite{kgcf06}, 
         HD139664 \cite{kgcf06}).
}
\label{fig:asymm}
\end{figure}

While to first order debris disks are rings of material, even if
the location and breadth of the rings is wavelength dependent,
on closer inspection those that have been imaged 
with sufficient clarity also exhibit significant asymmetries.
Different types of asymmetries have been identified which can be grouped
into the following categories: warps, spirals, offsets, brightness asymmetries,
and clumps.
The observed structures are summarised in Fig.~\ref{fig:asymm} and are
discussed in more detail below.

\subsubsection{Warps}
\label{sss:warps}
A warp\index{warp} arises when the plane of symmetry of a disk varies 
with
distance from the star.
It is only edge-on extended debris disks for which a warp
can be seen, since this orientation allows the plane of
symmetry at any given distance to be readily identified with the location
of the maximum surface brightness there.
Both of the edge-on disks with significant extension, $\beta$ Pictoris
and AU Mic, are warped \cite{hllc00,liu04}, as is the structure of the
zodiacal cloud in the solar system \cite{wdtf99}.
Recent observations of $\beta$ Pictoris suggest that
its warp may in fact not be continuous, and that there are two
separate disks with different planes of symmetry \cite{goli06}.
Since the images which show warps in debris disks have been made in
scattered light, it is important to point out that care must be
(and has been) taken when interpreting these observations, since
asymmetric scattering (i.e., the effect that causes back scattering to be stronger 
than forward scattering) can introduce perceived asymmetries into
observations of an otherwise axisymmetric disk \cite{kj95}.
A warp has also been identified in the face-on disk of TW Hydra \cite{rwm05}
from analysis of the emission spectrum which is affected by the fact that the warp
prevents light from the star reaching the outer portions of the disk.
This method of detecting a warp was possible for TW Hydra which has a
classical T Tauri disk, but is not possible for face-on debris disks which
are optically thin.

\subsubsection{Spirals}
\label{sss:spirals}
The disk of HD141569 is seen to be significantly extended with dust
out to 1200AU where there are two M stars of similar age which are
likely to be weakly bound to the star \cite{wrbz00}.
The radial distribution of dust is peaked at 150 and 250 AU.
Optical coronagraphic imaging shows that both of these rings is a
tightly wound spiral \cite{clam03}.
The diffuse emission from 300-1200 AU also forms a more open
spiral structure, with one, possibly two arms.
Two armed open spiral structure is also reported in younger transition disks,
such as AB Aur \cite{lolh06}.
Recent observations of Vega suggest that its extended sub-mm emission 
is condensed into two spirals (Holland et al., in prep.).

\subsubsection{Offsets}
\label{sss:offsets}
The star is not always at the centre of the rings.
This effect was first predicted \cite{wdtf99}, but then later
dramatically seen in optical images of the Fomalhaut disk \cite{kgc05}.
The Fomalhaut disk is narrow, and its proximity of 7.8 pc allowed the radius
to be measured with great accuracy as a function of azimuth.
With a mean disk radius of 133 AU, an offset of 15 AU was measured
with significant confidence.
The centre of the $\epsilon$ Eridani disk is also seen to be offset \cite{ghwd05},
however the lower resolution of the sub-mm observations and
its more complicated clumpy structure make the interpretation of this
measurement less clear.
Nevertheless, it is interesting to note that for the cases where such
measurements can be made (nearby bright disks), an offset is seen.

\subsubsection{Brightness Asymmetries}
\label{sss:bright}
The offset effect was first predicted from observations of the HR4796
disk \cite{tele00}.
This edge-on disk was seen to be $\sim 5$\% brighter on the NE than the
SW side, an asymmetry which was attributed to an offset.
However, there are other interpretations of brightness asymmetries,
since all of the spiral, offset and clump structures could appear
as brightness asymmetries when seen edge-on.
In other words, this class is likely another manifestation of one of the
other types of structure.
Indeed the $\beta$ Pictoris structure now attributed to a clump (see below)
was originally seen as an asymmetry \cite{lp94}.
Likewise, the brightness asymmetry seen in mid- to far-IR images of
the Fomalhaut disk \cite{stap04}, and which gets stronger
at shorter wavelengths, can likely be attributed to the offset seen in optical
images \cite{kgc05}.
Other disks with brightness asymmetries include HD32297 \cite{kala05}
and HD15115 \cite{kfg07}, for which the asymmetries are particularly
pronounced.
The latter is an example of a \textit{needle} disk\index{needle disk}, 
which is seen to
extend to significantly larger distances on one side of the star than the
other.
It is not clear if this is a brightness asymmetry (and the shorter side
extends out to the same distance but at a level below the detection limit)
or whether the two sides really are truncated at different outer radii.

\subsubsection{Clumps}
\label{sss:clumps}
The most common type of asymmetry seen in debris disks is a change in
brightness with azimuth around the ring, with much of the emission
concentrated in one or more clumps.
The clearest example of this phenomenon is the $\epsilon$
Eridani disk which is a narrow ring at 60 AU with
a well resolved inner hole \cite{ghmj98}.
The sub-mm images show four clumps of varying brightness within this ring.
The interpretation of this structure has been confounded by the
ubiquity of background galaxies which appear randomly across sub-mm
images.
However, the rapid proper motion of this star, which is at 3.6pc, has allowed
non-moving background objects to be identified, with three of the clumps confirmed
as real using imaging covering a time-span of $\sim 5$ years
\cite{ghwd05}.
While the inner hole of the Vega disk is seen less clearly in
850 $\mu$m imaging, the emission in this disk, which
is being seen close to face-on, is concentrated in
two clumps that are equidistant from the star, but asymmetric in
brightness \cite{hgzw98}.
The clumps are confirmed in mm-wavelength interferometry \cite{kso01,whkh02},
but appear at different locations in 350 $\mu$m imaging
\cite{mdvg06}, and not at all in far-IR images \cite{su05}, although that may
be because of the low resolution of these observations.
Other disks with clumps include Fomalhaut \cite{hgdw03}, although
this may be a manifestation of the offset, $\beta$ Pictoris \cite{tele05},
for which a brightness asymmetry appears to be originate in a clump with
a sharp inner edge, and AU Mic \cite{liu04}, for which clumps are seen
at a range of offsets from the star (although note that given the interpretation
of the axisymmetric disk structure \cite{sc06}, all of these clumps are likely
to be at the same distance from the star, just seen in projection).

\subsubsection{No detectable asymmetry}
\label{sss:noasymm}
Some of the resolved disks from Table \ref{tab:disks} 
exhibit no discernible asymmetry in their structure.
These are $\tau$ Ceti \cite{gwhd04}, HD107146 \cite{ardi04},
HD181327 \cite{schn06}, HD53143 and HD139664 \cite{kgcf06},
and HD92945 \cite{goli07}.
However, this does not necessarily mean that the disks are symmetrical, 
since some of these images do not have the resolution and/or sensitivity
to detect even large scale asymmetries.

\section{Debris disk models}
\label{s:mod}
The observed radial distribution of dust in debris disks can be explained as a
consequence of planetesimal belt dynamics.
Here I build up a disk dynamical theory\index{disk dynamical theory}
 which explains how dust is created in a planetesimal
belt, and how the combination of gravity, collisional processes and radiation forces
conspire to make the radial distribution of dust vary as a function of grain
size.

\subsection{The planetesimal belt}
\label{ss:pb}
First it is assumed that the outcome of planet formation was to create a
ring of planetesimals at a radius $r$ and of width $dr$.
The dominant force acting on these planetesimals is the gravity of the
star, and all material within the belt orbits the star.
These orbits are defined by their semimajor axis, $a$, eccentricity, $e$,
and orbital inclination, $I$, along with three angles defining the orientation
of the orbit (longitude of pericentre, $\varpi$, and longitude of ascending
node, $\Omega$), and the position within it (e.g., mean longitude, $\lambda$,
or true anomaly, $f$).
There is a distribution of orbital elements which is assumed to be
independent of size for the largest planetesimals.
This is not the case during planet formation, wherein larger objects
grow rapidly specifically because they have lower eccentricities and
inclinations than smaller objects.

The size of planetesimals ranges from some maximum diameter $D_{\rm{max}}$
down to dust of size $D_{\rm{min}}$, and the size distribution is defined
by the amount of cross-sectional area $\sigma(D)dD$ in each size bin of width 
$dD$;
cross-sectional area is defined such that a spherical particle has an area
of $\sigma=\pi (D/2)^2$.
Taking the size distribution to be described by a power law,
\begin{equation}
  \sigma(D) \propto D^{2-3q},
  \label{eq:sigd}
\end{equation}
it follows that, as long as the index $q$ is in the range 5/3 to 2,
the total amount of cross-sectional area in the belt, $\sigma_{\rm{tot}}$,
is dominated by the smallest objects within it, whereas its mass is dominated by
the largest objects.

\subsection{Collisions}
\label{ss:coll}
While eccentricities and inclinations of planetesimals are assumed to be
small, the resulting relative velocities are large enough that collisions are 
destructive.
This is necessary if dust is to be produced in collisions rather than
lost in growth to larger sizes \cite{dd05}.

Within the planetesimal belt collisions between planetesimals of different
sizes are continually occurring.
The result of such collisions is that the planetesimals are broken
up into fragments with a range of sizes.
If the outcome of collisions is self-similar (i.e., the size distribution
of the fragments is scale invariant), and the range of sizes in the distribution
is infinite, then the resulting size distribution has an exponent with $q=11/6$
\cite{tin96}.
In this situation the planetesimal belt forms what is known as a collisional
cascade\index{collisional cascade},
and the size distribution remains constant, with mass flowing from
large planetesimals to small grains.

The outcome of a collision depends on the specific incident kinetic energy, $Q$.
Catastrophic collisions\index{catastrophic collisions}
 are defined as collisions in which the largest
fragment produced in the collision has less than half the mass of original object. 
In general particles are destroyed in collisions with similar sized particles.
In the strength regime, $D<150$m, the outcome of a collision is determined by the
strength of a planetesimal and the specific incident kinetic energy required to
destroy it, $Q_{\rm{D}}^\star$, decreases with size.
In the gravity regime, $D>150$m, the fragments created in the collision tend to
reassemble under the action of their own gravity, so that a larger input energy is needed
to catastrophically destroy a planetesimal, and in that regime $Q_{\rm{D}}^\star$
increases with size.

The mean time between collisions for dust in the size range which contributes
the majority of the total cross-sectional area in the collisional cascade
can be approximated by \cite{wdtf99}:
\begin{equation}
  t_{\rm{col}} = t_{\rm{per}}/4\pi \tau_{\rm{eff}},
  \label{eq:tcol}
\end{equation}
in years, where $t_{\rm{per}}=a^{1.5}M_\star^{-0.5}$ is the orbital period,
and $\tau_{\rm{eff}} = \sigma_{\rm{tot}}/(2\pi r dr)$ is the effective optical
depth of the belt, a (wavelength independent) geometrical quantity that could
also be called the surface density of cross-sectional area.

Equation (\ref{eq:tcol}) usually applies to the smallest dust grains in
the cascade.
Larger objects have much longer collisional lifetimes, since there is a
lower cross-sectional area in the cascade with sufficient incident energy
to induce catastrophic destruction.
Their lifetime scales $\propto D^{5-3q}$ (i.e., $\propto D^{-0.5}$ for a
canonical collisional cascade size distribution).
For details of how the planetesimal strength, $Q_{\rm{D}}^\star$, and orbital
eccentricity $e$ affect the collision lifetime the reader is referred to
\cite{wd02,wsgb07}.

\subsection{Radiation forces}
\label{ss:rf}
The orbits of small grains are affected by the interaction of the grains
with stellar radiation \cite{bls79}.
This is caused by the fact that grains remove energy from the radiation
field by absorption and scattering, and then re-radiate that energy
moving with the particle's velocity.
The resulting radiation force has two components: a radial
force, known as radiation pressure\index{radiation pressure},
and a tangential force, known as Poynting-Robertson drag
(P-R drag)\index{Poynting-Robertson (P-R) drag}.
The parameter $\beta$ is the ratio of the radiation force to that of
stellar gravity and is mostly a function of particle size (since both
forces fall off $\propto r^{-2}$)
\begin{equation}
  \beta = F_{\rm{rad}}/F_{\rm{grav}} = 7.65 \times 10^{-4} (\sigma/m)
  \langle Q_{\rm{pr}} \rangle _{T_\star} L_\star/M_\star,
\end{equation}
where $\sigma/m$ is the ratio of a particle's cross-sectional area to its
mass (in m$^2$ kg$^{-1}$), $Q_{\rm{pr}}$ depends on the optical properties
of the particle, and $L_\star$ and $M_\star$ are in solar units.
For large spherical particles $\beta = (1150/\rho D)L_\star/M_\star$,
where $\rho$ is the particle density in kg m$^{-3}$, and $D$ is in $\mu$m.
For smaller particles $\beta$ tends to a value which is independent
of size (see chapter by Li).

\subsubsection{Radiation pressure}
\label{sss:rp}
Radiation pressure essentially causes a particle to \textit{see} a smaller
mass star by a factor $(1-\beta)$.
It is immediately clear that particles with $\beta>1$ are not bound and leave
the system on hyperbolic trajectories.
However, the effect of radiation pressure is also seen at lower
values of $\beta$, since it means that particles orbiting at the same
semimajor axis have different orbital periods, since $t_{\rm{per}} = 
[a^3/M_\star(1-\beta)]^{0.5}$.

\begin{figure}
\centering
\includegraphics[height=5cm]{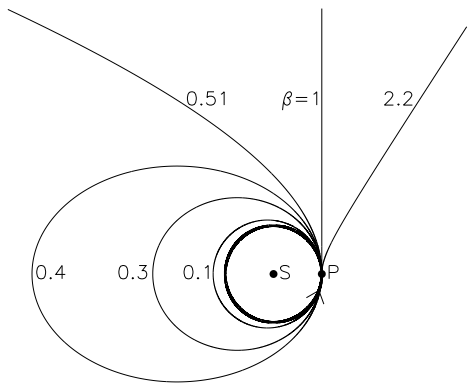}
\caption{Orbits of particles of different size (and so different $\beta$)
created in the destruction of a planetesimal originally on a circular orbit
\cite{wdtf99}.
The collision event occurs at point P.
Particles with $\beta>0.5$ are on unbound orbits.
}
\label{fig:rporbits}
\end{figure}

Most importantly, though, particles created in the destruction
of a parent planetesimal have a range of sizes and so $\beta$.
All particles start with the same position and velocity as the parent,
but have different orbital elements because they move in different
potentials.
For a parent with an orbit defined by $a$ and $e$ broken up at a true anomaly $f$,
the new orbital elements are
\begin{eqnarray}
  a_{\rm{new}} & = & a(1-\beta)[1-2\beta[1+e\cos{f}][1-e^2]^{-1} ]^{-1}, \\
  e_{\rm{new}} & = & [e^2 + 2\beta e \cos{f} + \beta^2]^{0.5}/(1-\beta)
\end{eqnarray}
(see Fig.~\ref{fig:rporbits}).
This means that, with a small dependence on where around the orbit the
collision occurs, it is particles with $\beta>0.5$ that are unbound and leave
the system on hyperbolic trajectories.
Since particles just above the radiation pressure blow-out 
limit\index{(radiation pressure) blow-out limit} survive much
longer than orbital timescales, this rapid loss causes a truncation in the
collisional cascade for small sizes below which $\beta>0.5$.

\subsubsection{P-R drag}
\label{ss:pr}
P-R drag causes dust grains to spiral into the star while at the same
time circularising their orbits (with no effect on the orbital plane).
For an initially circular orbit, this means that particles migrate in
from $a_1$ to $a_2$ on a timescale
\begin{equation}
  t_{\rm{pr}} = 400M_\star^{-1}(a_1^2-a_2^2)/\beta
  \label{eq:tpr}
\end{equation}
in years.
On their way in particles can be destroyed in collisions with other particles,
become trapped in resonance with planets \cite{djxg94}, pass through secular 
resonances, be scattered out of the system by those planets \cite{mm05}, or
be accreted onto the planets.
If none of these occurs, the particle sublimates close to the
star once its temperature reaches above $\sim 1500$ K.
This drag force is thus another potential loss mechanism
for dust from the collisional cascade.

It is evident that, since $t_{\rm{pr}} \propto D$ and $t_{\rm{col}} \propto D^{0.5}$, 
P-R drag can only be relevant for small particles.
Assuming that particles affected by P-R drag contribute little to the
total cross-sectional area, the particle size at which P-R drag becomes important 
can be estimated from equations (\ref{eq:tcol}) and (\ref{eq:tpr}),
\begin{equation}
  \beta > \beta_{\rm{pr}} = 5000 \tau_{\rm{eff}} (r/M_\star)^{0.5}.
\end{equation}
Since the smallest grains that may be influenced by P-R drag are those with
$\beta \approx 0.5$ it follows that P-R drag does not affect the evolution of
any grains in the disk if $\tau_{\rm{eff}} > 10^{-4}(r/M_\star)^{0.5}$,
as in this case all bound grains have collisional lifetimes that are shorter
than their P-R drag lifetimes.

\begin{figure}
\centering
\includegraphics[height=8cm]{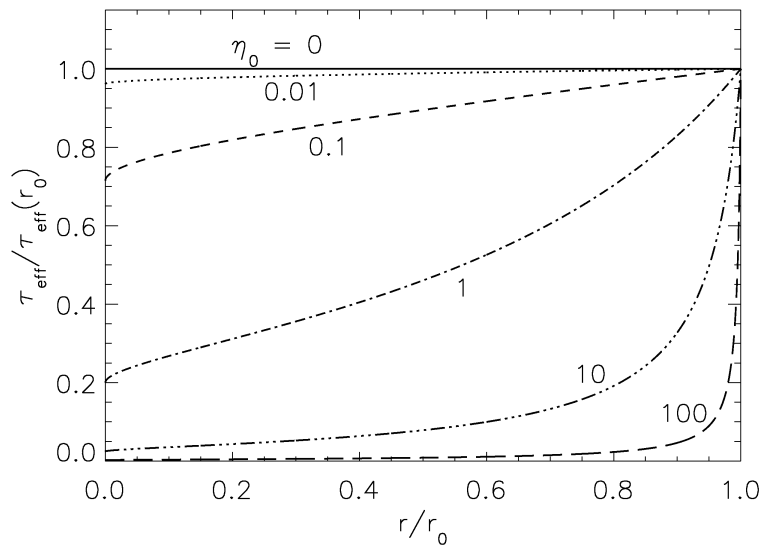}
\caption{Distribution of surface density for dust grains evolving from
a point of origin in a planetesimal belt at $r_0$ inwards due to
P-R drag while also being depleted due to mutual collisions \cite{wyat05a}.
}
\label{fig:prwcoll}
\end{figure}

This back-of-the-envelope calculation was demonstrated more quantitatively
in \cite{wyat05a} which considered the ideal case of a planetesimal belt which
produces grains all of the same size.
The spatial distribution of such grains as they evolve due to collisions and
P-R drag is given by
\begin{equation}
  \tau_{\rm{eff}}(r) = \tau_{\rm{eff}}(r_0)/[1+4\eta_0(1-\sqrt{r/r_0})],
\end{equation}
where $\eta_0=t_{\rm{pr}}/t_{\rm{col}}=
5000\tau_{\rm{eff}}(r_0)\sqrt{r_0/M_\star}\beta^{-1}$ and $r_0$ is the
radius of the planetesimal belt (see Fig.~\ref{fig:prwcoll}).
For $\eta_0 \ll 1$ the majority of the grains make it to the star without suffering
a collision, whereas for $\eta_0 \gg 1$ the grain population is significantly depleted
before the grains make it to the star and so are confined to the vicinity of the
planetesimal belt.
This model also illustrates how it is not possible to invoke P-R drag to create a
large dust population close to the star, since the maximum possible 
surface density of grains that reach the star in this model is
$5 \times 10^{-5} \beta M_\star^{0.5}r^{-0.5}$.

\subsection{Disk particle categories}
\label{ss:categories}
\begin{figure}
\centering
\includegraphics[height=4.5cm]{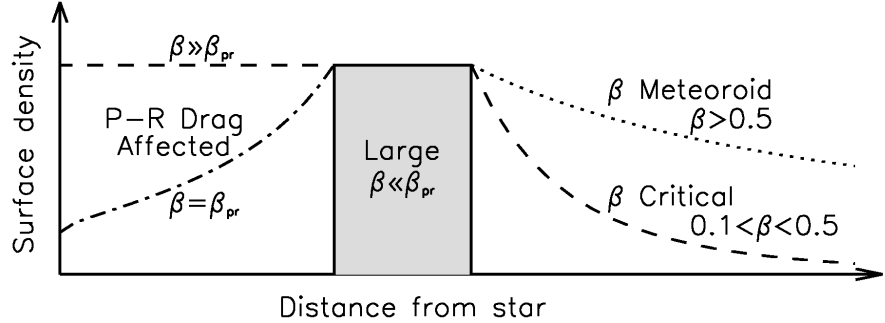}
\caption{Surface density distribution of particles created in a
planetesimal belt \cite{wyat99}.
Particles of different sizes have different $\beta$ and so have different
radial distributions.
The main categories are: large grains, which have the same distribution as
the planetesimals; $\beta$ critical and $\beta$ meteoroid grains, which extend
much further from the star; and P-R drag affected grains, which extend inwards
toward the star.
}
\label{fig:categories}
\end{figure}

The preceding discussion motivates the division of a debris disk
into distinct particle categories which is summarised in
Fig.~\ref{fig:categories}:
\begin{itemize}
  \item \textbf{Large grains} ($\beta \gg \beta_{\rm{pr}}$):
    these are unaffected by radiation forces and have the same spatial distribution
    as the planetesimals;
  \item \textbf{P-R drag affected grains}\index{P-R drag affected grains}
    ($\beta \approx \beta_{\rm{pr}}$):
    these are depleted by collisions before reaching the star;
  \item \textbf{P-R drag affected grains} ($\beta_{\rm{pr}} < \beta < 0.5$):
    these are largely unaffected by collisions and evaporate on reaching the star;
  \item \textbf{$\beta$ critical grains}\index{$\beta$ critical grains}
    ($0.1 < \beta < 0.5$):
    these are on bound orbits and while the inner edge of their distribution follows
    that of the planetesimals, the outer edge extends out to much larger distances;
  \item \textbf{$\beta$ meteoroid grains}\index{$\beta$ meteoroid grains}
    ($\beta > 0.5$):
    these are blown out on hyperbolic orbits as soon as they are created.
\end{itemize}

The presence of different categories in a disk depends on the density of the
planetesimal belt.
Broadly speaking, the large, $\beta$ critical and $\beta$ meteoroid categories
are always present (even if the quantities of the latter two relative to the
large grain population are not well known).
Thus there are two main types of disk: dense disks that are dominated
by collisions which have few P-R drag affected grains, and tenuous disks that
are dominated by P-R drag in which P-R drag affected grains are present.

\subsubsection{Collision versus P-R drag dominated disks}
\label{sss:colprdom}
The majority of the debris disks that can be detected at present are
squarely in the collision dominated\index{collision dominated}
regime, since $\eta_0 \gg 1$ \cite{wyat05a}.
In the absence of P-R drag the spatial distribution of material becomes very simple.
It is even possible to make the simplifying assumption that the $\beta$
meteoroid population is negligible, because such grains are lost on
timescales that are short compared with even the shortest lifetimes in
the large grain population.
Further ignoring complications due to the eccentricities of the $\beta$
critical grains, the disk can be modelled as material entirely
constrained to the planetesimal belt with a size distribution that extends 
in a single power law (eq.~\ref{eq:sigd}) down to the blow-out limit
\cite{wd02}.
While clearly a simplification, this model of the disk is far better than
one in which it is comprised of grains all of the same size,
since it acknowledges that the
dust we see has to originate somewhere.
Numerical simulations have also been performed to determine the
size and spatial distribution in the collision dominated limit in more detail
\cite{tab03,ta07}, and further analytical quantification of the distributions
in this limit is also possible \cite{sc06}.

The high sensitivity of Spitzer means that more recently relatively
low density disks have been detected for which $\eta_0$ is as low as 1 meaning
that P-R drag is expected to sculpt the inner edges of these disks \cite{wssr07}.
The effect of P-R drag also needs to be accounted for when studying dust in the
solar system, since $\eta_0 \approx 2 \times 10^{-3}$ in the asteroid belt.
It is important to emphasise this point, since it means that the dynamics of
dust in the zodiacal cloud is fundamentally different to that of extrasolar systems,
albeit in an understandable way.
It is also becoming clear that, while stellar wind forces\index{stellar 
wind forces} are relatively weak in
the solar system providing a drag force $\sim 1/3$ that of P-R drag (see
chapter by Mann), such 
forces may be important for other stars.
While the mass loss rates, $dM_{\rm{wind}}/dt$, of main sequence stars are poorly known, 
it is thought that the low luminosity of M stars means that this force may be responsible
for the dearth of disks found around late type stars \cite{pjl05}.
Since stellar wind forces act in a similar manner to P-R drag, they can be
accounted for in the models by reducing $\eta_0$ by a factor of
$[1+(dM_{\rm{wind}}/dt)c^2/L_\star]$ \cite{jura04,mkw07}.

Thus, while it is usually the case that the collision dominated approximation
is most appropriate, models which describe the distribution of material evolving under the 
action of collisions and drag forces continue to be of interest.
While a study which takes into account the full range of sizes in the disk
has yet to be undertaken, it is possible to see that since grains are typically
destroyed in collisions with similar sized objects, the outcome of such a model will
be similar to assuming that grains of different sizes have spatial distributions that
can be characterised by different $\eta_0$, with large grains having high $\eta_0$ 
and small grains having low $\eta_0$.
This means that the size distribution would be expected to vary significantly with
distance from the star.

\subsection{Comparison with observations}
\label{ss:comp}
This model has had considerable success at explaining the observed radial structure
of debris disks.
For example, using the collision dominated
assumption with the dust confined to the planetesimal belt provides an adequate fit to the
emission spectrum of disks like that of Fomalhaut \cite{wd02} for which the
radius of its planetesimal belt is well known \cite{hgdw03}.
It can also explain the structures of the disks which are seen to be
considerably extended and to exhibit a gradient in grain size throughout the disk
(AU Mic and $\beta$ Pictoris).
These observations are explained as $\beta$ critical dust being created in
planetesimal belts which are closer to the star \cite{anlp01,ab05,sc06}.
Further, the emission spectrum of the TWA7 disk is consistent with the 
distribution of
dust expected from inward migration from the planetesimal belt by stellar wind drag
\cite{mkw07}.

Thus these studies show that the observed dust distributions can be successfully
explained within the framework of a realistic physical model.
Such a model is an absolute requirement if any asymmetries seen in the disk structure
are to be interpreted correctly, since even the axisymmetric dust distribution is
different to that of the planetesimals, which hints that its asymmetric distribution
may also differ.
On a more basic level, this shows that the location of the dust in a debris disk
does not necessarily directly pinpoint the location of the planetesimals.

Despite the successes of the disk dynamical theory it is important to point out that
it is not yet a predictive theory.
There are too many uncertainties regarding the expected size distribution
at very small sizes (e.g., because it depends on the size distribution created in
collisions), and regarding the optical properties of those grains and the magnitude
of stellar wind drag, to predict how bright a disk known from far-IR
measurements (of its large grain population) will appear in scattered light images (which are 
sensitive to the $\beta$ critical and $\beta$ meteoroid grains).

\section{Interaction between planets and planetesimal belt}
\label{s:plpl}
Consider now one modification to the planetesimal belt model described in
\S \ref{s:mod}, which is that there is a planet orbiting in this system.
The gravitational perturbations of that planet will affect the orbits of
both the planetesimals and the dust.
It turns out that these perturbations are predicted to cause exactly
the same set of features as observed in debris disks
(Fig.~\ref{fig:asymm}).

The planet's perturbations can be broken down, both mathematically and
conceptually, into three types: secular, resonant and short period
perturbations.
For a detailed description of this dynamics the reader is referred to Murray
\& Dermott (1999) \cite{md99}.
Its secular perturbations\index{secular perturbations}
 are the long term consequence of having the planet
in the disk, and these perturbations are equivalent to the perturbations
from the wire that would be obtained by spreading the mass of the planet
around its orbit with a density in accordance with its velocity at each
point;
they affect all material in the disk to some extent.
Its resonant perturbations\index{resonant perturbations}
are the forces which act at specific radial locations
in the disk where planetesimals would be orbiting the star with a period that
is a ratio of two integers times that of the planet.
At such locations the planetesimal receives periodic kicks from the planet
which can make such locations either extremely stable or extremely unstable.
All other perturbations are short period and can be assumed
to average out on long enough timescales, although they are responsible for
important processes such as scattering of planetesimals.

\subsection{Secular perturbations}
\label{ss:secpert}
To first order a planet's secular perturbations can be separated into two components,
one arising from the eccentricity of its orbit, and the other from its inclination.
For high eccentricities and inclinations these two components are linked,
and here I only consider the low eccentricity and inclination case.

\subsubsection{Planet eccentricity: spirals and offsets}
\label{sss:spe}
The consequence of the planet's eccentricity is to impose an eccentricity
onto the orbits of all planetesimals in the disk.
It does this in such a way that a planetesimal's eccentricity vector,
defined by $z=e \times \exp{i\varpi}$, precesses around a circle centred
on the forced eccentricity\index{forced eccentricity} vector, 
$z_{\rm{f}}$; i.e.,
\begin{equation}
  z(t) = z_{\rm{f}} + z_{\rm{p}}(t),
\end{equation}
where the forced eccentricity is set by a combination of the
planet's eccentricity and the ratio of the planetesimal and
planet semimajor axes
\begin{equation}
  z_{\rm{f}} = [b_{3/2}^2(\alpha_{\rm{pl}})/b_{3/2}^1(\alpha_{\rm{pl}})]z_{\rm{pl}},
\end{equation}
and the proper eccentricity\index{proper eccentricity} precesses around
a circle the radius of which is determined by the initial conditions
\begin{equation}
  z_{\rm{p}}(t) = e_{\rm{p}} \times \exp{i(At+\beta_0)},
\end{equation}
at a fixed rate given by
\begin{equation}
  A = 0.25n (M_{\rm{pl}}/M_\star)\alpha_{\rm{pl}}\bar{\alpha}_{\rm{pl}}b_{3/2}^1(\alpha_{\rm{pl}}).
\end{equation}
In the above equations, $\alpha_{\rm{pl}}=a_{\rm{pl}}/a$ and $\bar{\alpha}_{\rm{pl}}=a/a_{\rm{pl}}$
for $a_{\rm{pl}}<a$ and $\alpha_{\rm{pl}}=\bar{\alpha}_{\rm{pl}}=a/a_{\rm{pl}}$ for $a_{\rm{pl}}>a$,
and $b_{3/2}^s(\alpha_{\rm{pl}})$ are the Laplace coefficients. 
These equations have been given for the case of a system with
one planet.
However, the same decomposition into forced and proper elements is also
true in a system with multiple planets, except that the equations for the
forced eccentricity and precession rate $A$ involve sums over all planet properties
\cite{wdtf99}.

\begin{figure}
\centering
\begin{tabular}{c}
  \hspace{+0.8in} \includegraphics[height=7.0cm]{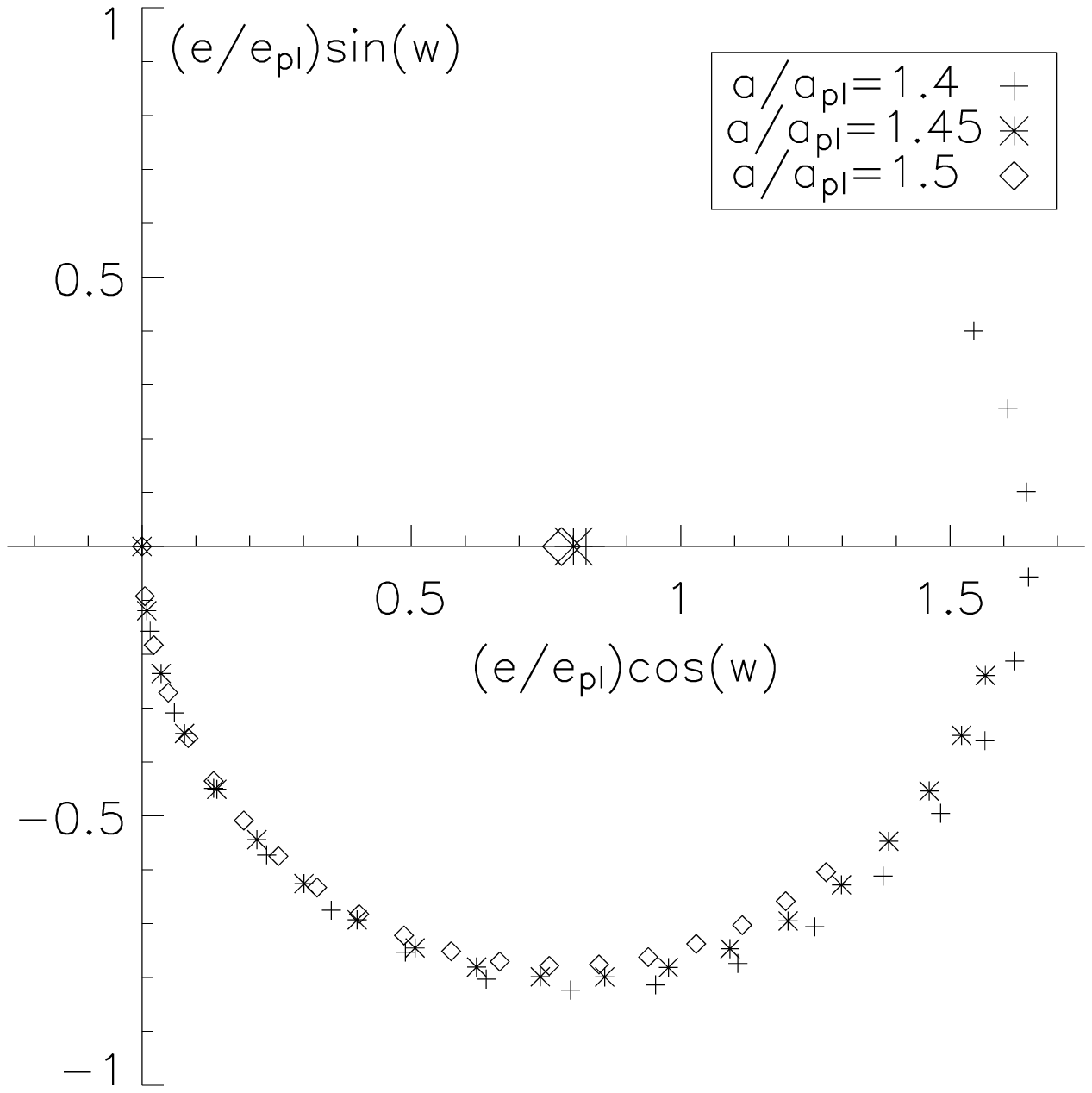} \\
  \hspace{-0.35in} \includegraphics[height=6.5cm]{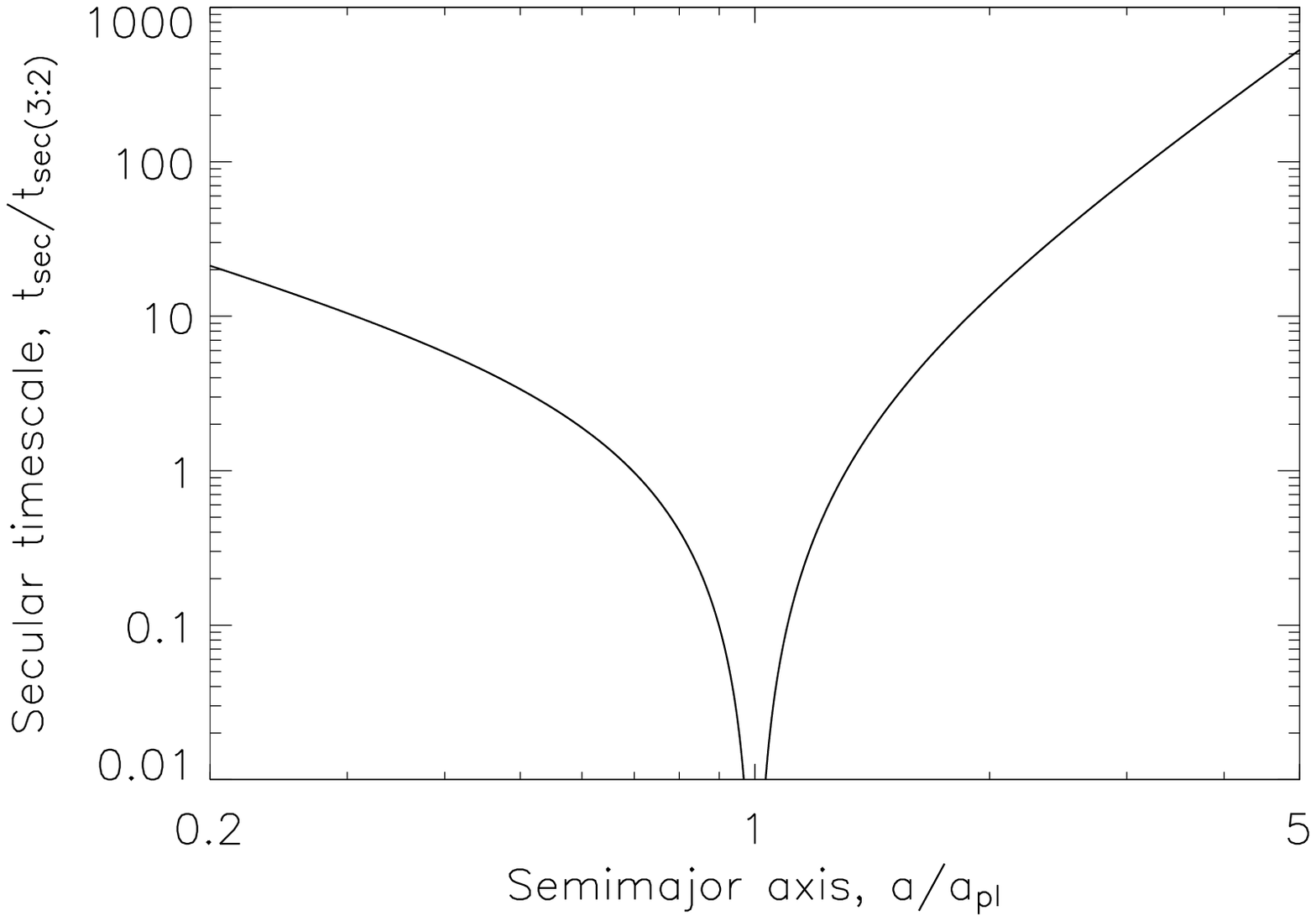}
\end{tabular}
\caption{Effect of the secular perturbations of an eccentric planet on
planetesimal orbits \cite{wyat05b}.
\textbf{Top} Evolution of the eccentricity vectors of planetesimals at
1.4, 1.45 and $1.5a_{\rm{pl}}$.
All vectors start at the origin (circular orbits) and precess around
the forced eccentricity imposed on them by the planet.
The symbols are plotted at equal timesteps.
\textbf{Bottom} Precession rate for planetesimals at different distances
from the  planet.
These are given as the timescale to complete one precession ($2\pi/A$)
relative to that timescale for planetesimals at $a=1.31a_{\rm{pl}}$
which is given by $0.651 a_{\rm{pl}}^{1.5} M_\star^{0.5}/M_{\rm{pl}}$.
}
\label{fig:speevol}
\end{figure}

The evolution of a planetesimal's eccentricity vector is illustrated in
Fig.~\ref{fig:speevol}, which shows how the orbits of planetesimals at
1.4, 1.45 and 1.5 times semimajor axis of the planet evolve if they
start on initially circular orbits.
This is equivalent to a situation which might arise following the formation
of a planet on an eccentric orbit, since the planetesimals would have formed
on roughly circular orbits.
As well as a small change in forced eccentricity for planetesimals at
different distances from the planet, their precession rates are substantially
different.
This means that the dynamical structure of an extended planetesimal disk evolves
following the formation of the planet:
planetesimals close to the planet which have completed several precessions can
be considered to have eccentricity vectors evenly spread around circles
centred on the forced eccentricity, while those further away have pericentre
orientations and eccentricities which change with distance from the star.

\begin{figure}
\centering
\begin{tabular}{c}
  \includegraphics[height=5.5cm]{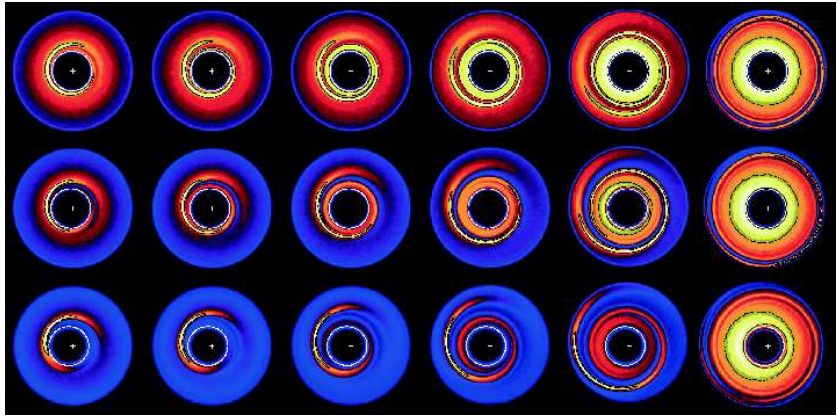}
\end{tabular}
\caption{Spatial distribution of planetesimals affected by the secular perturbations
of an eccentric planet: spiral structure propagating outward through an 
extended planetesimal belt outside a planet \cite{wyat05b}.
The subpanels show, from left to right, snapshots of the disk at times 0.1,0.3,1,3,10,100 times the 
secular precession timescale at $1.31a_{\rm{pl}}$ since the perturbing planet was introduced
into the disk.
From top to bottom the panel show the impact of planets with an eccentricity of 0.05,0.1 and 
0.15.
}
\label{fig:speevolspiral}
\end{figure}

The resulting dynamical structure can be readily translated into a spatial
distribution by creating a model in which planetesimals are distributed
randomly in longitude, $\lambda$, and with other orbital elements taken from
appropriate distributions for the given time.
This is shown in Fig.~\ref{fig:speevolspiral} for the planetesimals outside the
planet's orbit.
It is seen that the planetesimals exhibit spiral structure which propagates away
from the planet.
A similar spiral structure is formed in the planetesimal disk interior to the planet, again 
propagating away from the planet with time.

It is possible that the effect seen in Fig.~\ref{fig:speevolspiral} may be the explanation
for the tightly wound spiral structure seen at 325 and 200 AU in the HD141569 disk \cite{clam03}.
Since the rate at which the spiral propagates away from the planet is determined by
the planet's mass, this means that the observed structure allows the planet's mass to
be estimated, assuming the time since the planet formed can also be estimated.
For HD141569 this results in the putative planet having a mass greater than that of Saturn,
given the 5 Myr age of the star.
A tightly wound spiral structure is also seen in Saturn's rings \cite{cpdb05}
which is explained by a similar model in which the secular perturbations
arise from the oblateness of the planet (rather than from an eccentric perturber), and
the ring material is assumed to have formed in a relatively recent event (rather than
with the planet).

\begin{figure}
\centering
\begin{tabular}{c}
  \includegraphics[height=5.5cm]{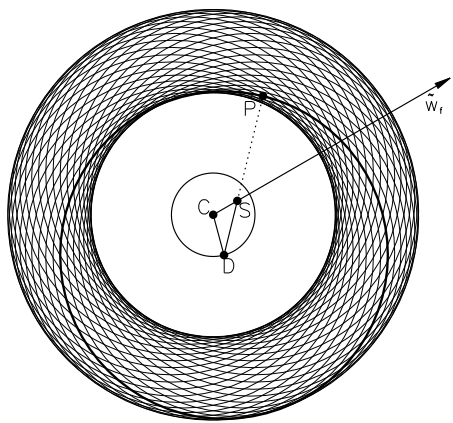}
\end{tabular}
\caption{Spatial distribution of planetesimals affected by the secular perturbations
of an eccentric planet:
offset structure imposed at late times on planetesimals 
all at the same semimajor
axis $a$ \cite{wdtf99}.
The planetesimals form a uniform torus around the star at $S$, but one which has its centre
at a point $D$ which is offset from the star by a distance $ae_{\rm{f}}$ in the direction of
the forced apocentre.
}
\label{fig:speevolspiral2}
\end{figure}

At late times, when the material at the same semimajor axis has eccentricity vectors that
are evenly distributed around circles, the resulting disk no longer exhibits spiral structure,
but it does exhibit an offset.
This is illustrated in Fig.~\ref{fig:speevolspiral2}, 
although it is also apparent in Fig.~\ref{fig:speevolspiral} at late times.
The offset is proportional to the forced eccentricity imposed on the planetesimals
(and so proportional to the planet's eccentricity), with material on the side of
the forced pericentre being closer to the star than that on the side of the forced apocentre.

This offset was originally predicted from a brightness asymmetry in the HR4796 
disk \cite{wdtf99,tele00}, and was called \textit{pericentre 
glow}\index{pericentre glow} because the asymmetry was
thought to arise from the material on the pericentre side being hotter than that on the
apocentre side.
It was found that the observed brightness asymmetry could have been caused
by a planet with an eccentricity as small as 0.02, demonstrating that even
moderate planet eccentricities can have observable signatures.
However, little information is available from this structure about the mass of the planet, except
that it must be sufficiently massive for the pericentres to have been randomised given the
age of the star.
For HR4796, this means that its putative planet would have to have a mass $>10M_\oplus$,
although the interpretation of this asymmetry is complicated by the stellar mass
binary companion to HR4796A, the orbit of which is unknown at present, but which 
could also be responsible for an offset of the required magnitude.
Nevertheless, an offset has been seen directly in the structure of the Fomalhaut disk
\cite{kgc05}.
This star also has a common proper motion companion \cite{bshb97}, but this is too
distant to be responsible for an offset of the observed magnitude.

\subsubsection{Planet inclination: warps}
\label{sss:spi}
The consequence of secular perturbations caused by the planet's inclination is directly
analogous to the consequence of its eccentricity (\S \ref{sss:spe}), except that in this
case it is the planetesimal's inclination vector, $y=I \times \exp{i\Omega}$,
which precesses around a forced inclination\index{forced inclination}.
The precession rate is also the same, except that it is reversed in sign (i.e.,
the inclination vector precesses clockwise on a figure analogous to Fig.~\ref{fig:speevol}).
In a system with just one planet the forced inclination vector is simply the
orbital plane of the planet ($y_{\rm{f}} = I_{\rm{pl}} \times \exp{i\Omega_{\rm{pl}}}$).
Since the choice of the zero inclination plane is arbitrary it can be set to be
the planet's orbital plane ($y_{\rm{f}} = 0$) making it easy to see that at late times,
planetesimals at the same semimajor axis will have orbital planes distributed randomly about
the orbital plane of the planet.
However at early times, should the initial orbital plane of the planetesimals be different
to that of the planet at say $y_{\rm{init}}$, then the situation will be that
material close to the planet will be distributed randomly about the planet's orbital
plane $y_{\rm{pl}}$, while that far from the planet will still be on the original
orbital plane $y_{\rm{init}}$.
A smooth transition between the two occurs at a distance from the star
which depends on the mass of the planet and the time since the planet formed,
much in the same way that spiral structure propagates away from the planet.

This has been proposed as the explanation of the warp in the $\beta$ Pictoris 
disk, since other lines of evidence have pointed to a Jupiter mass planet
at $\sim 10$ AU \cite{rsss94, bm00}, and given the age of the star $\sim 12$ Myr, it 
is reasonable to assume that a warp would be seen at $\sim 80$ AU at the current 
epoch (if the planet formed very early on).
Many observations of the disk, including the warp and the radial distribution
(see \S \ref{ss:axi}), can be explained with such a model \cite{anlp01},
although it would be worth revisiting this model in the light of
the observations which showed the warp is less of a smooth transition between
two orbital planes and looks more like two distinct disks \cite{goli06}.

This mechanism does not just produce a warp in a young disk.
As long as there are two or more planets in the system on different orbital
planes a warp would also be seen at late times, once all the
planetesimals have precessed so that their distribution is symmetrical
about the forced inclination plane, since multiple planets would mean
that the forced inclination plane varies with distance from the
star (e.g., it is aligned with each of the planet's orbital planes
at the semimajor axes of those planets).
The zodiacal cloud in the solar system is an example of a warped
old disk \cite{wdtf99}.

\subsection{Resonant perturbations}
\label{ss:respert}
Mean motion resonances\index{mean motion resonances} are locations at 
which planetesimals orbit the
star an integer $p$ times for every integer $p+q$ times that the planet
orbits the star.
The nominal location of a resonance is at a semimajor axis of
\begin{equation}
  a_{(p+q):p} = a_{\rm{pl}}(1+q/p)^{2/3}.
  \label{eq:ares}
\end{equation}
Planetesimals at a range in semimajor axis about this
nominal value may be trapped in resonance, but not all of
those in this range are necessarily in the resonance.

\subsubsection{Resonant geometry}
\label{sss:resgeom}
\begin{figure}
\centering
\begin{tabular}{cccc}
  \includegraphics[height=2.85cm]{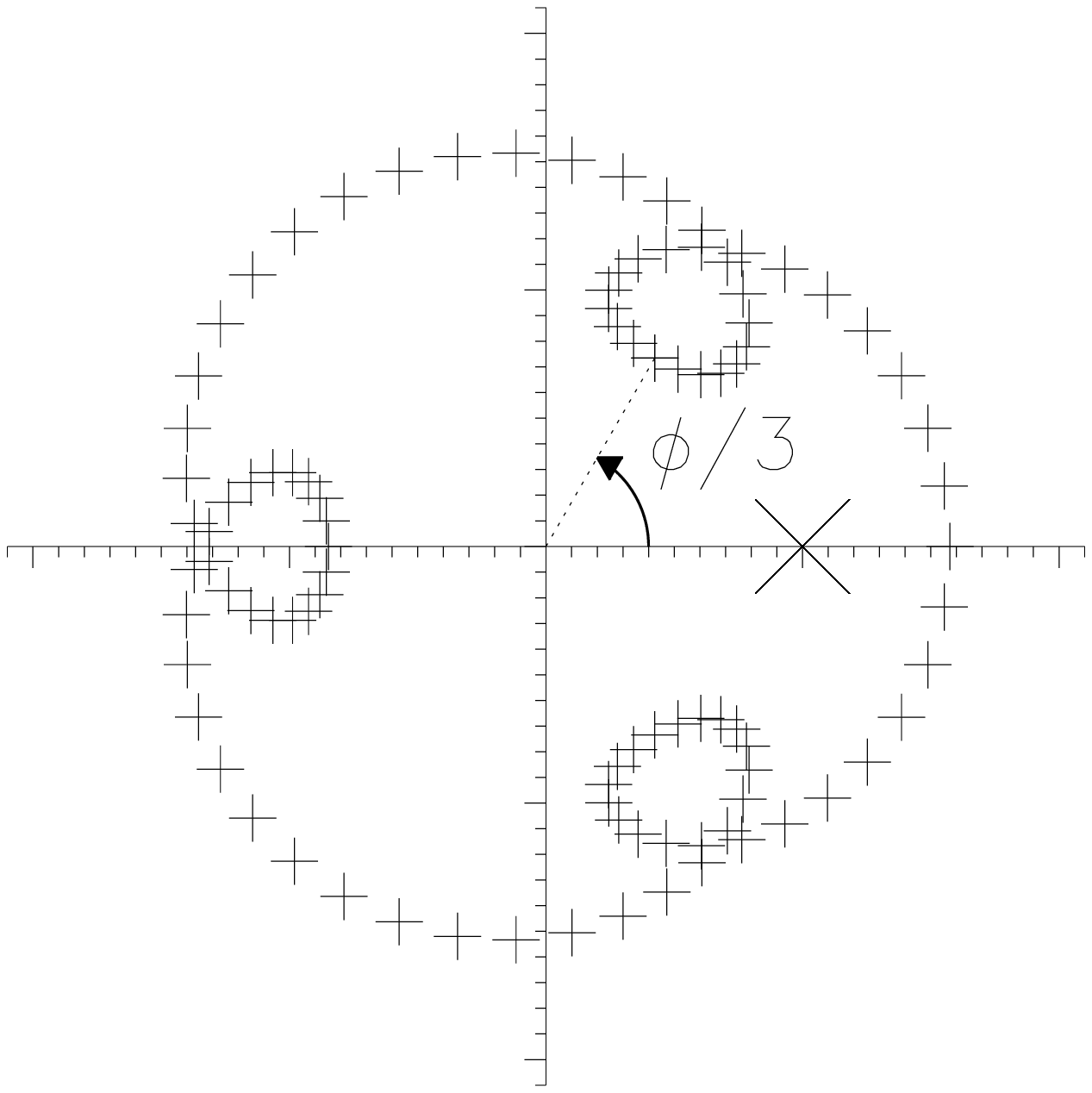} &
  \includegraphics[height=2.85cm]{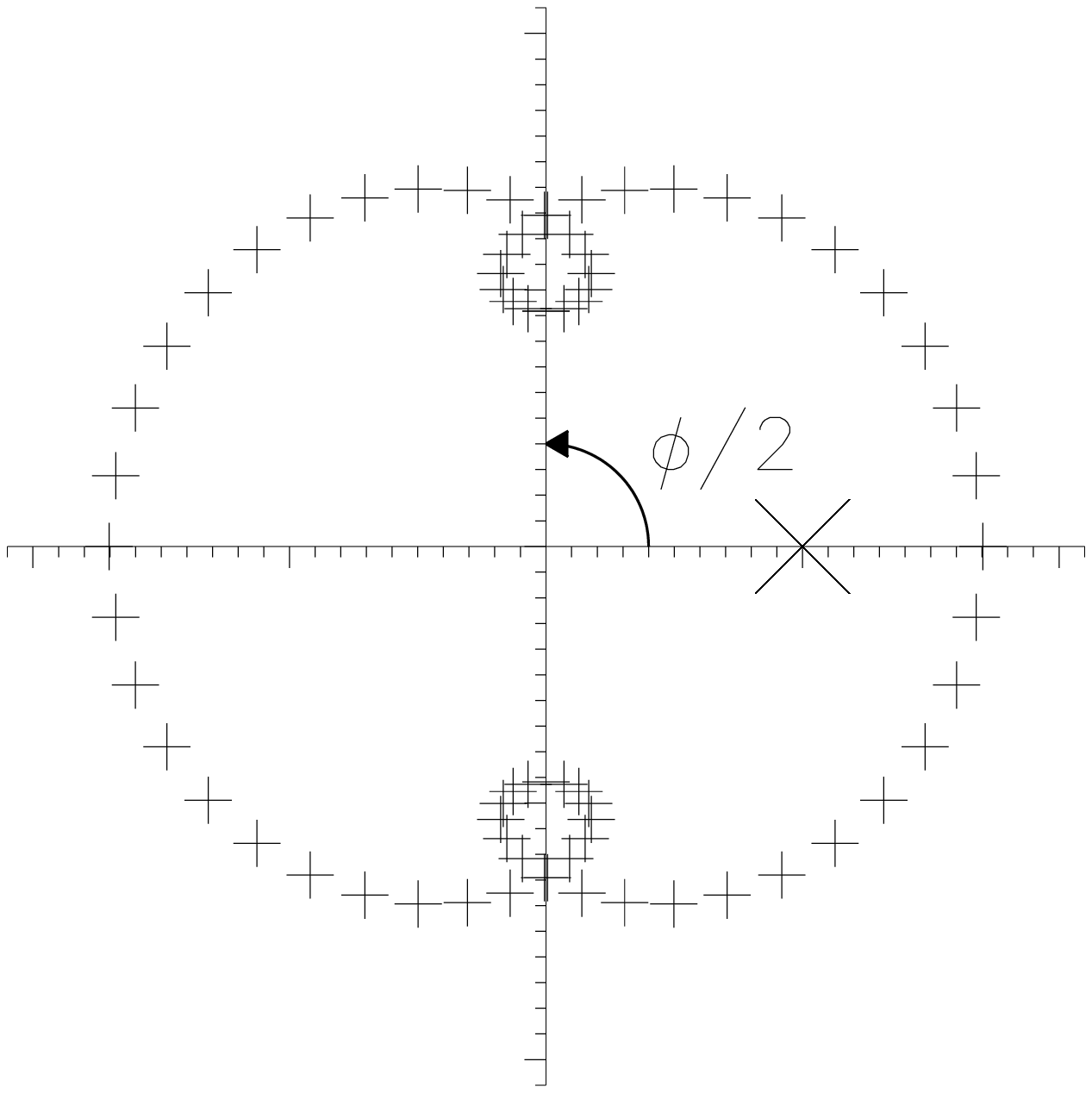} &
  \includegraphics[height=2.85cm]{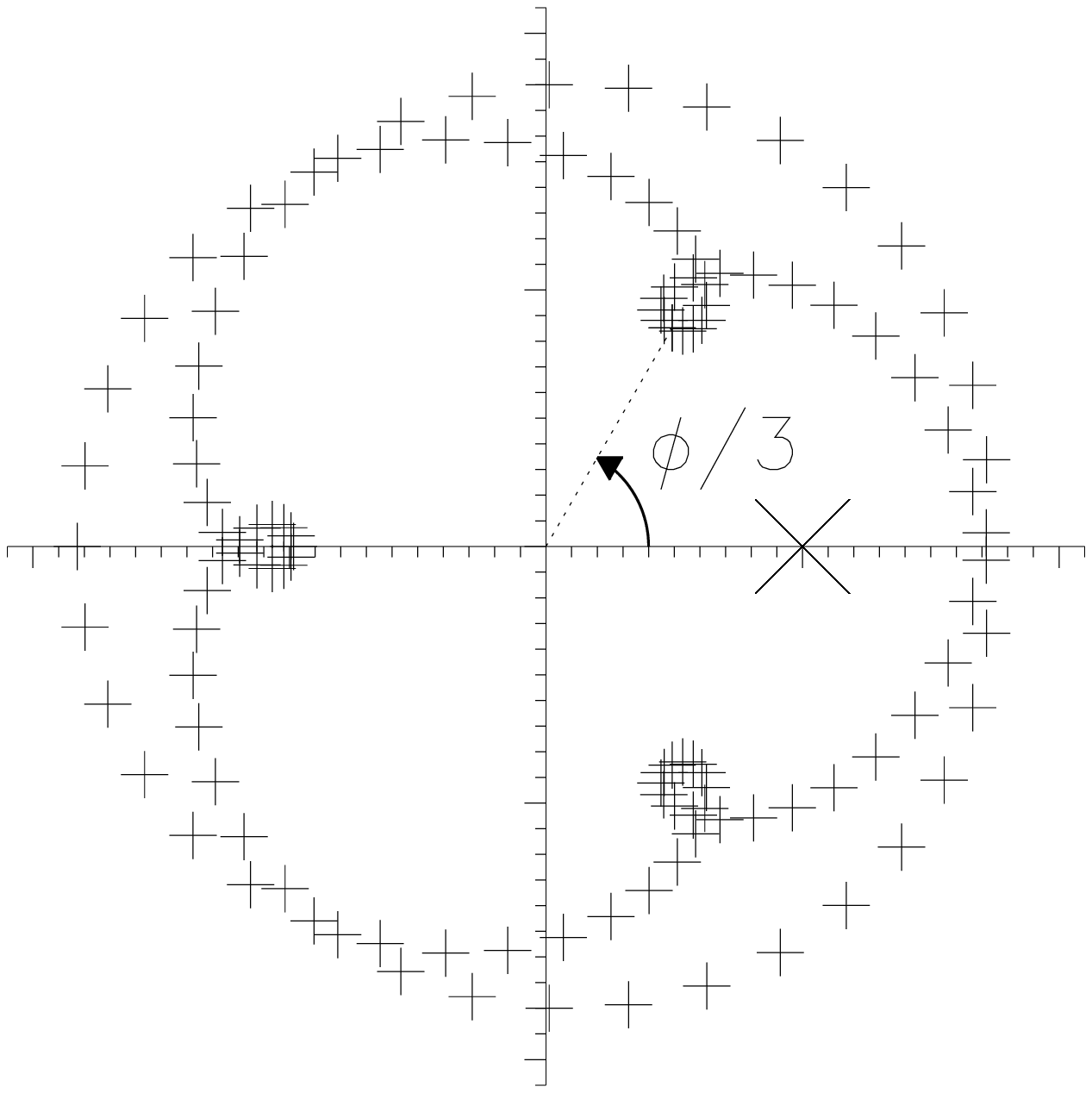} &
  \includegraphics[height=2.85cm]{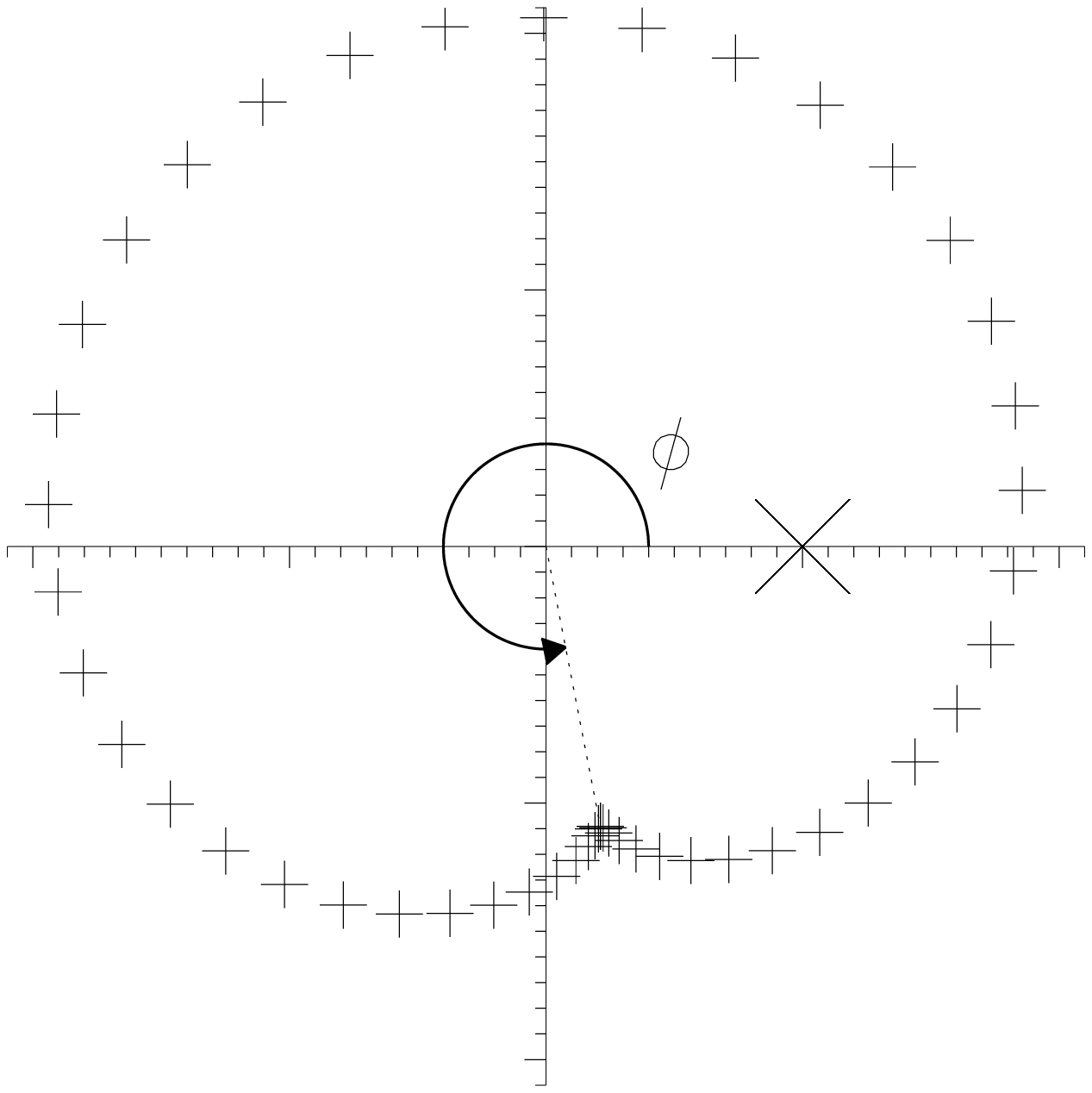}
\end{tabular}
\caption{Path of resonant orbits in the frame co-rotating with a planet \cite{wyat03}.
On all panels the planet, located at the cross, is on a circular orbit,
while the planetesimals' orbits have an eccentricity of 0.3.
The planetesimals are plotted with a plus at equal timesteps through their
orbit, each point separated by 1/24 of the planet's orbital period.
The resonances shown are from left to right, with increasing distance from the planet,
the 4:3, 3:2, 5:3 and 2:1 resonances.
}
\label{fig:res}
\end{figure}

The importance of a resonance can be understood purely from geometrical
reasons. 
Fig.~\ref{fig:res} shows the path of planetesimals on resonant orbits
in the frame co-rotating with the planet.
The pattern repeats itself so that the planetesimal
always has a conjunction with the planet (i.e., the two are at the same
longitude) at the same point in its orbit for $q=1$ resonances, or at the
same two points in its orbit for $q=2$ resonances.
This is important because the perturbations to the planetesimal's orbit
are dominated by those at conjunction which means that the planetesimal receives
periodic kicks to its orbit from the planet which are always in the same
direction (if the orbit is unchanged by those perturbations).
This is not quite true for the $p=1$ resonances, since the cumulative effect
of the perturbations around the orbit are also relevant in this case.

The resonant geometry discussed in the preceding paragraph can be used to infer
the loopy patterns on a figure such as Fig.~\ref{fig:res}, but it does not
specify the location of the planet with respect to those loops.
That is specified by the planetesimal's resonant argument\index{resonant 
argument}
\begin{equation}
  \phi = (p+q) \lambda - p \lambda_{\rm{pl}} - q \varpi.
  \label{eq:phi}
\end{equation}
The resonant argument is important, since the ratio $\phi/p$ is the relative longitude
of the planet when the planetesimal is at pericentre, an angle which is noted
on Fig.~\ref{fig:res}; i.e., it determines where along the planetesimal's orbit it
receives kicks from the planet's gravity.

The same combination of angles occurs in the planet's disturbing function, and the
forces associated with a resonance are those involving the relevant resonant
argument \cite{md99}.
A planetesimal is said to be in resonance if its resonant argument is librating
about a mean value (e.g., a sinusoidal oscillation), rather than circulating (e.g.,
a monotonic increase or decrease).
The mean value about which the resonant argument librates is typically
$180^\circ$, since in this configuration it can be shown that the resonant
forces impart no angular momentum to the planetesimal.
However, in some instances asymmetric libration\index{asymmetric 
libration}
 occurs, where $\langle \phi 
\rangle \ne 180^\circ$, because the equilibrium solution requires
resonant forces to impart angular momentum to the planetesimal (see section
on resonant trapping and \S \ref{ss:resringpr}).
Asymmetric libration also occurs for the $p=1$ resonances 
(e.g., the 2:1 resonance), because in this configuration
angular momentum imparted to a planetesimal at conjunction is
balanced by the cumulative effect of the resonant forces around
the rest of the orbit \cite{mc05}.

\subsubsection{Resonant trapping}\index{resonant trapping}
\label{sss:restrap}
While resonances have non-zero width in semimajor axis, their width is finite and
they only cover a narrow region of parameter space.
Thus if a planet was introduced into an extended planetesimal belt, while
planetesimals at suitable semimajor axes might end up trapped in
resonance, such planetesimals would be relatively few.
However, resonances can be filled by either the planet or the planetesimals'
semimajor axes undergoing a slow migration, since when a planetesimal encounters
a planet's resonances the resulting forces can cause the planetesimal to become
trapped in the resonance.
Resonant forces could then either halt the planetesimal's migration, or make
it migrate with the planet, thus ensuring that the planetesimal maintains the
resonant configuration.
For example, it is thought that Pluto and most of the other Kuiper belt objects
that are in resonance with Neptune attained their resonant orbits during an
epoch when Neptune's orbit expanded following its formation \cite{malh95}.
There are a number of mechanisms which can be invoked to cause planets to migrate
outward, one of which is angular momentum exchange caused by scattering of
planetesimals \cite{hm99}, and another is interaction with a
massive gas disk \cite{mlpw07}.

The question of whether a planetesimal becomes trapped once it encounters a
planet's resonances is determined by two main factors: the mass of the planet
and the rate at which the planet or planetesimals are migrating.
For example, the probability of a low eccentricity
planetesimal being trapped into any given resonance with a planet migrating
on a circular orbit is determined by two parameters,
$\mu = M_{\rm{pl}}/M_\star$ and $\theta = \dot{a}_{\rm{pl}}/\sqrt{M_\star/a}$
\cite{wyat03}.
It is expected that the eccentricities of the planet and planetesimal orbits
would also affect the trapping probability.
Another important factor in determining which resonances are filled by
planet migration is the initial distribution of the planetesimals with
respect to the planet, since this determines how many planetesimals would 
encounter a given resonance in the course of the migration, given that some
may already have been trapped in another resonance.
The dominant resonances in the Kuiper belt are the 4:3, 3:2, 5:3 and 2:1 
resonances, which can partly be explained by the fact that first order resonances 
(i.e., $q=1$ resonances) are stronger than second order resonances (i.e., $q=2$ 
resonances), and so on.

Resonances are important not only for a disk's dynamical structure, but also for
the spatial distribution of material, since Fig.~\ref{fig:res} illustrates how
planetesimals that are all in the same resonance that have similar resonant
arguments would tend to congregate at specific longitudes relative to the 
planet.
That is, while any one planetesimal is on an elliptical orbit, that orbit
would take it in and out of regions of high planetesimal density, i.e., clumps.
These clumps would appear to be orbiting the star with the planet.
The number of clumps formed by resonances is given by $p$ (e.g., Fig.~\ref{fig:res}).
An important factor in the formation of resonant clumps is the planetesimals'
eccentricities, since the clumps only become pronounced at high eccentricities;
resonant planetesimals on circular orbits have an axisymmetric distribution.
Once trapped resonant forces can excite the planetesimals so that they 
become eccentric.

\begin{figure}
\centering
\begin{tabular}{c}
  \includegraphics[height=7.0cm]{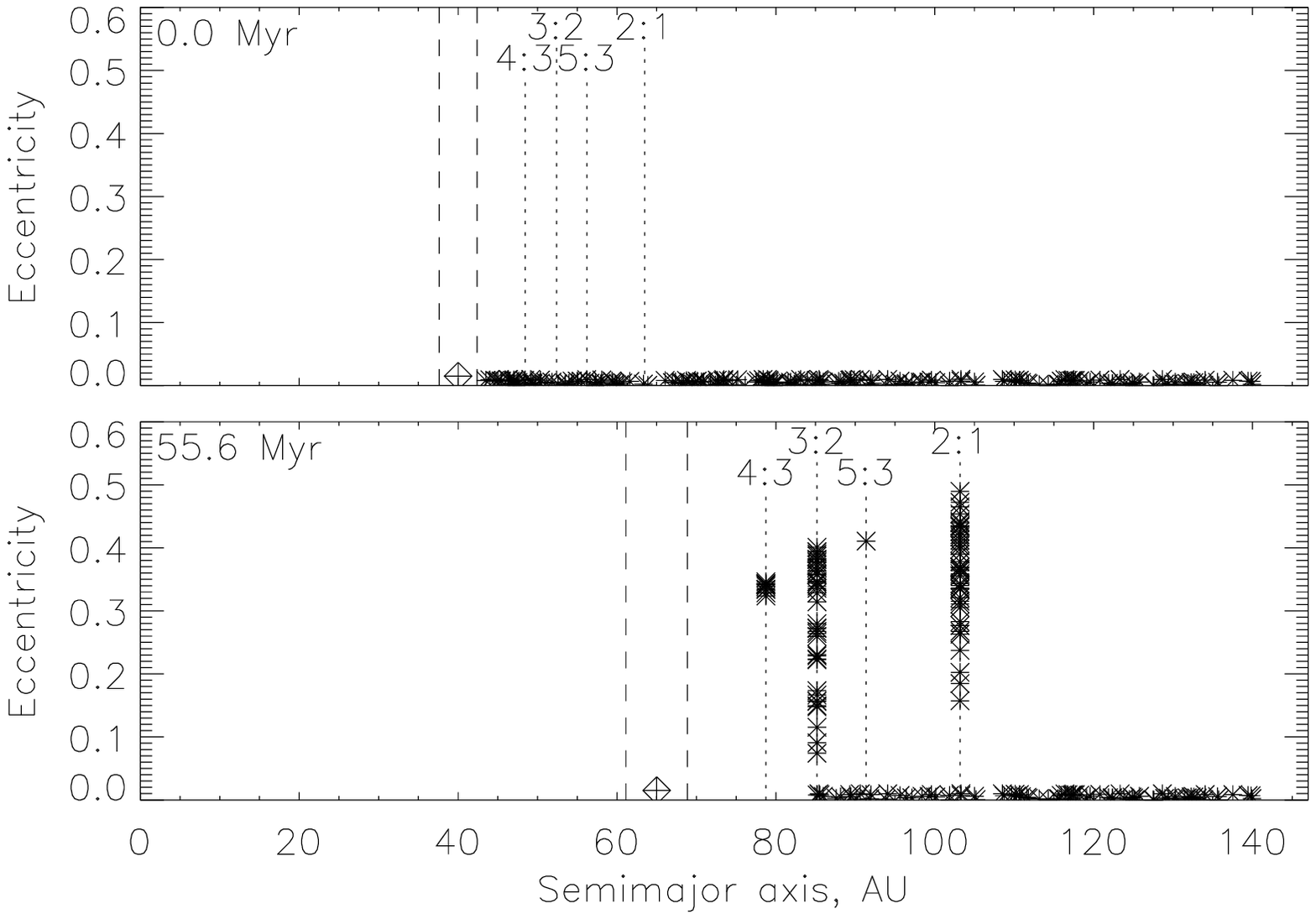} \\[0.2in]
  \includegraphics[height=7.6cm]{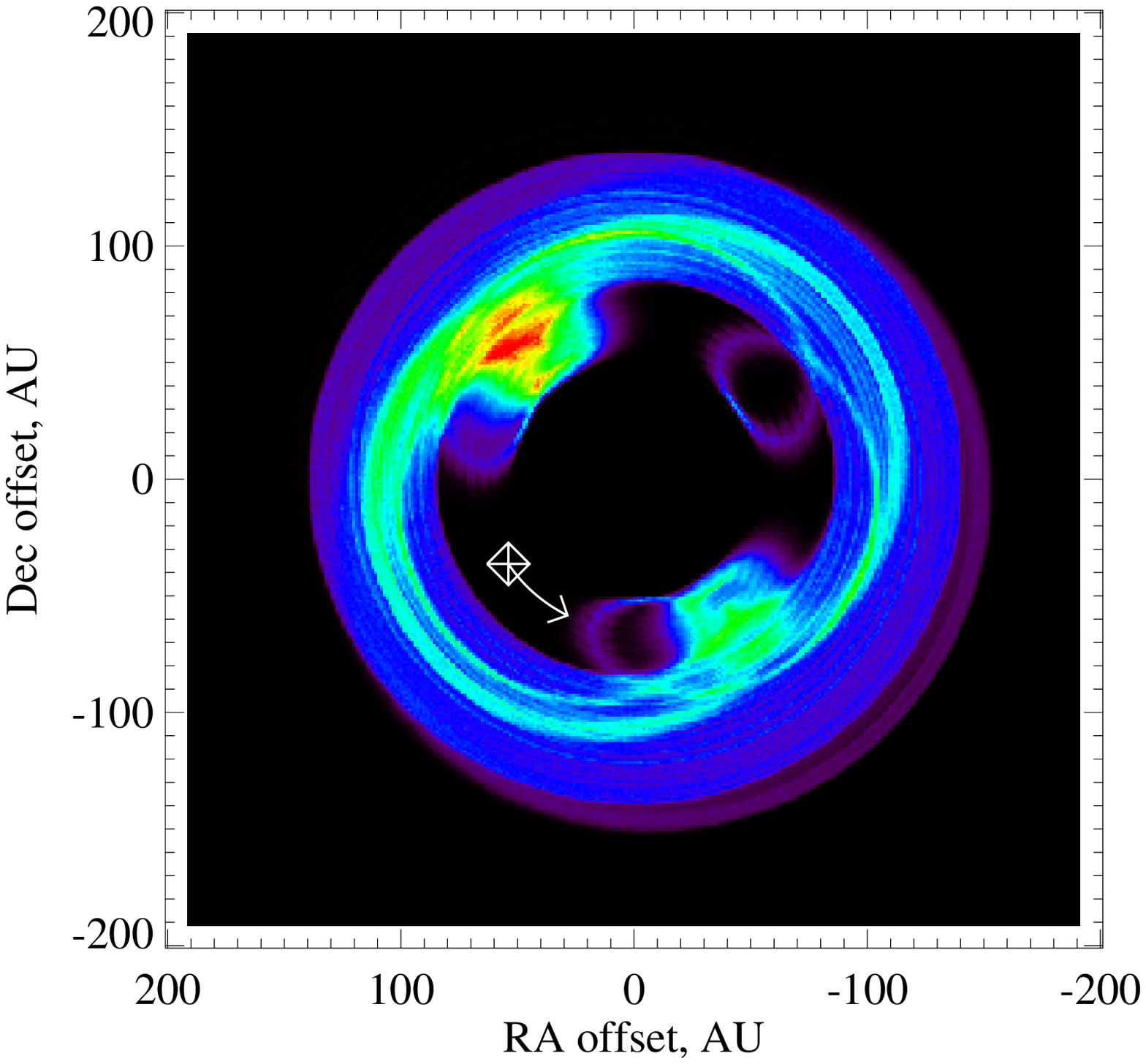}
\end{tabular}
\caption{Structure imposed on an initially axisymmetric planetesimal disk
by the outward migration of a planet \cite{wyat03}.
This model was proposed to explain the clumpy structure of the Vega disk
and involves a Neptune mass planet which migrated from 40-65 AU
over 56 Myr.
\textbf{Top} Dynamical structure of the planetesimal disk, eccentricity versus
semimajor axis, at the beginning and end of the migration.
The planet is shown with a diamond, and the location of its resonances with dotted
lines.
The chaotic region of resonance overlap\index{resonance overlap}
 is shown with dashed lines.
\textbf{Bottom} Spatial distribution (surface density) of planetesimals at the
end of the migration.
The planet is shown with a diamond and the arrow shows its direction of orbital
motion.
}
\label{fig:vegamod}
\end{figure}

This mechanism was invoked to explain the clumpy structure of Vega's debris
disk (Fig.~\ref{fig:vegamod}; \cite{wyat03}).
In that model two clumps form in the planetesimal distribution because of the
migration of a Neptune mass planet from 40 to 65 AU over 56 Myr.
As suggested by Fig.~\ref{fig:res} the clumps are the result of trapping
of planetesimals into the 3:2 resonance, with an asymmetry caused by planetesimals
in the 2:1 resonance.
The planet's mass and migration rate are constrained within the model, but not
uniquely, since it did not consider the origin of the planet's migration;
e.g., the same structure would arise from a 3 Jupiter mass planet
which completed the same migration over 3 Myr.
To break this degeneracy models would be required which cause
both planet migration and resonant trapping simultaneously, a task
which requires significant computing power.
Nevertheless, this model shows that observed structures have the potential
to tell us not only about the planets in a system, but also about that system's
evolutionary history.

The model has made predictions for:
\textbf{(i)} the location of the planet (none has been found at the level of
$<3M_{\rm{Jupiter}}$, \cite{mhw03});
\textbf{(ii)} the orbital motion of the clumps which should be detectable on
decade timescales (these observations will be made in the coming year, and in the
meantime marginal detection of orbital motion has been found in the clumpy structure
of the $\epsilon$ Eridani disk, \cite{pgc06});
\textbf{(iii)} lower level structure associated with the 4:3 and 5:3 resonances
(which may have been seen in 350 $\mu$m observations of the disk \cite{mdvg06}).

\subsubsection{Resonance overlap} 
\label{sss:resolap}
So far I have discussed the stabilising properties of resonances.
However, resonances can also be destabilising.
As mentioned above, resonances have finite width, and the $q=1$ resonances are 
strongest.
There is a region nearby the planet where its first order resonances overlap,
and planetesimals in such a region have chaotic orbits and are rapidly
ejected \cite{wisd80}.
The width of this region is
\begin{equation}
  |a-a_{\rm{pl}}|/a_{\rm{pl}} = 1.3(M_{\rm{pl}}/M_\star)^{2/7}.
\end{equation}
Instabilities of this type have been invoked to explain the cleared inner regions of
debris disks \cite{fq07}, as well as to estimate the location of a planet
inside an imaged planetesimal belt \cite{quil06}.
The same planet also imposes eccentricities on the planetesimals at the edge of
the resonance overlap region, and the magnitude of those eccentricities is dependent on the
mass of the planet, with more massive planets imposing larger eccentricities.
Since those eccentricities result in a sloping inner edge, it is also possible
to use the sharpness of the inner edge of a dust ring to set constraints on the
mass of the planet.
In this way the sharp inner edge of the Fomalhaut ring was used to determine that
its planet must be less massive than Saturn \cite{quil06}.

\section{Interaction between planets and dust}
\label{s:pldust}
The preceding section (\S \ref{s:plpl}) dealt specifically with the
structures imposed by planets on the planetesimal distribution.
This is an important first step, since the dust we see is derived from
those planetesimals.
However, as discussed in \S \ref{s:mod}, dust dynamics 
can be significantly different to that of the planetesimals, and so
it is not obvious to what extent the dust distribution will
follow that of the planetesimals.
A model which takes into account the production in collisions
of dust with a range of sizes and the subsequent dynamical evolution
of that material is usually beyond the scope of current computing
(and analytical) capabilities.
Such models will likely become more common-place as more detailed
observations demonstrate that more sophisticated models are
necessary to explain the observations.
For now, the types of structure which dust
dynamics would produce can be understood by considering the dynamical
evolution of dust grains released from a given planetesimal distribution.
Those grains might then be considered to evolve in the absence of
collisions, or in an idealised situation where the only
collisions which matter are those with grains of similar size
(and so which all have the same spatial distribution).

Here I consider the effect of dust grain dynamics on the structures
seen in a planetesimal belt in which some of the planetesimals are
in resonance with a planet (\S \ref{ss:dustmigpl}), and
the structures caused by trapping of dust into planetary
resonances (\S \ref{ss:resringpr}).

\subsection{Dust produced from resonant planetesimals}
\label{ss:dustmigpl}
Only the largest dust grains released from a resonant planetesimal
remain in that resonance, and the reason is the effect of radiation
pressure.
First of all, consider the orbits of grains which remain bound to
the star ($\beta<0.5$).
Radiation pressure has two effects:
it changes the semimajor axis of the dust grain (so that
$a^{'} = a[1-\beta]/[1-2\beta]$ for initially circular orbits),
and also the location of the resonance (the resonance for dust
is at a lower semimajor axis by a factor $(1-\beta)^{1/3}$
than given in equation \ref{eq:ares}).
This means that the larger a particle's $\beta$ (which typically
means the smaller its size, although see \S \ref{ss:rf} and chapter
by Li) the further it starts from the resonance, and while resonant forces 
can accommodate this offset by increasing the libration width for large 
grains, there comes a size at which particles are no longer in resonance.
Numerical simulations showed that for the 3:2 resonance the critical
size is that for which $\beta > \beta_{\rm{crit}}$, where
\begin{equation}
  \beta_{\rm{crit}} = 2 \times 10^{-3} (M_{\rm{pl}}/M_\star)^{0.5},
  \label{eq:betacrit}
\end{equation}
with a similar threshold for the 2:1 resonance \cite{wyat06}.
Since the geometry of Fig.~\ref{fig:res} is no longer valid for
non-resonant grains, such grains have an axisymmetric distribution.

Next, consider the orbits of dust grains which are released onto
hyperbolic orbits (i.e., $\beta>0.5$).
For $\beta=1$ dust no force acts on the grains, and such grains leave the system with a 
constant velocity (that of the orbital motion of the parent planetesimal) which rapidly 
approaches radial motion.
Assuming that these grains are created at a constant rate, this corresponds
to a surface density distribution which falls off $\propto r^{-1}$.
Since no force acts on the grains, one might naively expect no asymmetry
in their distribution.
However, their distribution can be non-axisymmetric if they are not produced
from an axisymmetric distribution of parent bodies.
Collision rates are highest between resonant planetesimals when they are in
the clumps, and this means that a greater fraction of the $\beta>0.5$ grains
created from planetesimal collisions have trajectories which originate in the
clumps.
The distribution of such grains should thus exhibit spiral structure which
emanates from the clumps (since while the motion of the dust is nearly radial,
the source region, the clumps, are in orbital motion around the star). 
However, not all $\beta>0.5$ grains are created in collisions between planetesimals
with a clumpy resonant distribution;
some originate in collisions between non-resonant grains with
$\beta_{\rm{crit}}<\beta<0.5$, and so would have an axisymmetric distribution.

\begin{figure}
\centering
\begin{tabular}{c}
  \includegraphics[height=9.2cm]{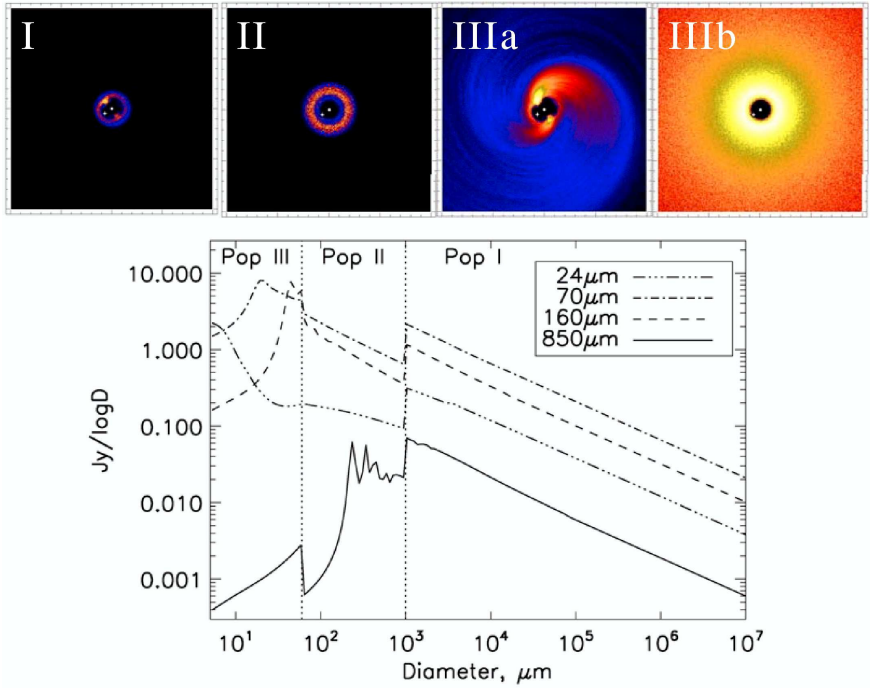}
\end{tabular}
\caption{Prediction for the structure of Vega's debris disk \cite{wyat06}.
\textbf{(Top)} Spatial distribution of dust in different
populations:
\textbf{(I)} large grains,
\textbf{(II)} intermediate grains,
\textbf{(IIIa)} small grains (created from large grains),
\textbf{(IIIb)} small grains (created from intermediate grains).
All panels cover the same region ($\pm 100$ arcsec from the star
which is shown by an asterisk at the centre);
the location of the planet is shown with a plus.
\textbf{(Bottom)} Contribution of different grain
sizes (and so different populations) to observations in different
wavebands.
The y axis is flux per log particle diameter, so that the area under the
curve is the total flux, and the relative contribution of different grain
sizes to that flux is evident from the appropriate region.
For the size distribution shown here the mid- to far-IR wavelength observations
are dominated by population III grains, while sub-mm observations are dominated
by population I grains.}
\label{fig:vegamoddust}
\end{figure}

This motivates a division of the dust produced in a resonant planetesimal disk
into four populations with distinct spatial distributions:
\textbf{(I)} large grains $\beta<\beta_{\rm{crit}}$ with a clumpy distribution,
\textbf{(II)} intermediate grains $\beta_{\rm{crit}}< \beta < 0.5$, with an axisymmetric
  distribution,
\textbf{(IIIa)} small grains $\beta>0.5$ from population \textbf{(I)} particles with extended
spiral structure,
\textbf{(IIIb)} small grains $\beta>0.5$ from population \textbf{(II)} particles with 
extended axisymmetric structure.
These distributions have been worked out numerically for the model presented in 
Fig.~\ref{fig:vegamod}, and the structures expected for the four populations
are shown on Fig.~\ref{fig:vegamoddust}.

Aside from ascertaining the distribution of different grain sizes, it is important
to determine which grain sizes actually contribute to the observation in question.
This chapter will not deal specifically with such issues, for which a knowledge of the
optical properties of the particles is needed, and for which the 
the reader is referred to the chapter by Li in this book.
However, the type of result that is obtained with such an analysis is illustrated
in Fig.~\ref{fig:vegamoddust}.
This shows how observations in different wavebands are sensitive to different grain
sizes and to different grain populations, with the shortest wavelengths probing
the smallest grains;
i.e., the disk would be expected to look different when observed at different wavelengths.
For the Vega disk, this is indeed seen to be the case \cite{mdvg06},
although this does not mean the dynamics of this disk is completely understood,
since the prediction for the spiral structure at the shortest wavelengths \cite{wyat06}
has yet to be confirmed, and the large observed mass loss rate remains to be explained
\cite{su05}.

The fact that disk structure is expected to be (and is seen to be) a strong function of both 
grain size and wavelength of observation is good because it means that multiple 
wavelength observations of the same disk provide a means to test different models for the 
origin of structure formation.
However, it also means that the models are becoming more complicated, and this means that
the interpretation of observed structure is no longer straight forward, since there
are multiple physical processes that have to be accounted for.
For example, it should also be noted that the model described above only took account of the 
effect of radiation pressure on the dust orbits, and the relative velocity imparted to 
collisional fragments may also be important \cite{kqls07}.

\subsection{Resonant trapping of dust by P-R drag}
\label{ss:resringpr}
Planetary resonances can also sculpt a dust disk even if the parent
planetesimals are not trapped in resonance, since the drag forces
which act on dust to make it migrate inwards (see \S \ref{ss:pr})
mean that the dust may have the opportunity to encounter a planet's resonances.
Resonant forces can then halt the migration causing a concentration of
dust along the planet's orbit known as a \textit{resonant
ring}\index{resonant ring}.
For the same geometrical reasons as outlined in \S \ref{sss:resgeom}, these
resonant rings are clumpy.

\begin{figure}
\centering
\begin{tabular}{c}
  \includegraphics[height=8cm]{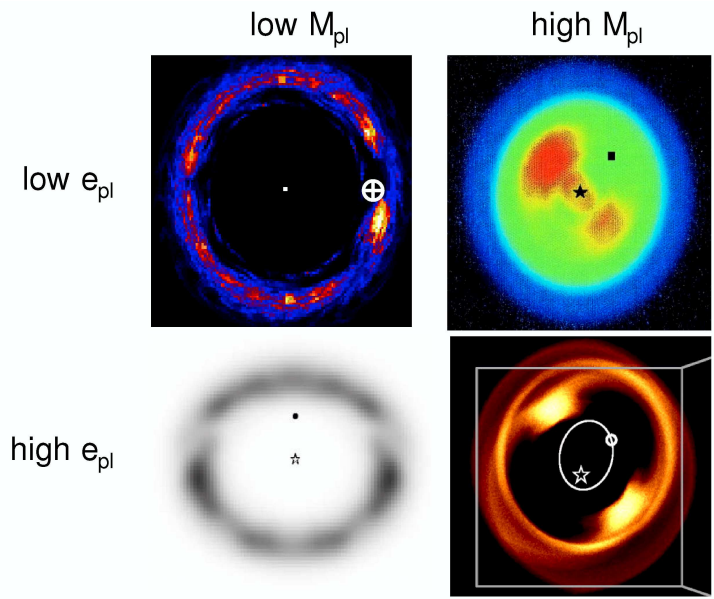}
\end{tabular}
\caption{Spatial distribution of dust which has migrated into the resonance
of a planet forming a \textit{resonant ring}.
The structure of the ring depends on the mass and eccentricity of the planet
\cite{kh03}, and examples of the four types of structure are taken from 
published models:
  low $M_{\rm{pl}}$, low $e_{\rm{pl}}$
    (model for the Earth's resonant ring \cite{djxg94});
  high $M_{\rm{pl}}$, low $e_{\rm{pl}}$
    (model for the Vega dust ring \cite{ogmt00});
  low $M_{\rm{pl}}$, high $e_{\rm{pl}}$
    (model for the $\epsilon$ Eridani dust ring \cite{qt02});
  high $M_{\rm{pl}}$, high $e_{\rm{pl}}$
    (model for the Vega dust ring \cite{whkh02}).
}
\label{fig:resrings}
\end{figure}

There are some important subtle differences in the structure of this type
of resonant ring compared with the resonant planetesimal rings.
One of these is the fact that the libration of $\phi$ is offset from $180^\circ$ so that
resonant forces can impart angular momentum to the particles to counteract that
lost by P-R drag.
This means that the loopy patterns on Fig.~\ref{fig:res}
are not symmetrical about the planet in such a way that the loop which is
immediately behind the planet is closer to the planet than that in front of it.
The magnitude of this effect is dependent on particle size ($\beta$).
The concentration of all the loops from the different resonances and particle
sizes behind the planet causes a clump to follow the planet around its orbit.
This is sometimes referred to as a trailing wake\index{trailing wake}.
This effect was responsible for the discovery of the first resonant ring,
since the zodiacal cloud was found to always be brighter in the direction behind 
the Earth's motion than in front of it \cite{djxg94}.
This was interpreted as dust trapped in $q=1$ resonances close to the
Earth (i.e., with $p>3$) (see top left panel of Fig.~\ref{fig:resrings}).
The structure of the Earth's trailing wake will soon be known in great
detail, as the infrared satellite Spitzer is currently flying directly through
the middle of it.
There is no evidence for a resonant ring associated with Mars \cite{krb00}, but
recent evidence shows that Venus has a resonant ring \cite{lm07}.

The structure of a resonant ring depends on the mass of the planet,
because the resonant forces from a more massive planet are stronger
meaning that dust can be trapped into resonances that are further
from the planet, e.g., the 3:2 and $p=1$ resonance such as the 2:1
and 3:1 resonances (see top right panel of Fig.~\ref{fig:resrings}).
The ring structure is also dependent on the planet's eccentricity
\cite{kh03} (bottom panels of Fig.~\ref{fig:resrings}).
However, one of the most important factors which determines that structure is
the spatial distribution of source planetesimals and the size distribution of
particles encountering the different resonances,
since it is that which determines which resonances are populated.
It is not easy to ascertain the expected structure of a resonant ring,
since a complete resonant ring model would have to consider the competition between
production and destruction in collisions and removal by P-R drag, on top
of which some fraction of the particles are trapped in different resonances
for varying durations.
Needless to say, current models make some approximations, and 
typically ignore collisions and consider only a relatively narrow range of
particle sizes that are assumed to evolve independently \cite{mm02,dm05,mwm05}.

One important point to consider is that for a resonant ring to form in this
way the dust must migrate inwards on a timescale that is shorter than the
timescale for it to be destroyed in collisions.
As discussed in \S \ref{sss:colprdom}, the role of P-R drag in affecting the
orbits of dust in the disks which are known about at present is negligible, since
the collision timescale is short.
Thus, while stellar wind drag forces may increase the drag rate for late type
stars, the expectation is that resonant rings of this type are not present
in the known disks \cite{wyat05a,kqls07}.
This serves as a caution that it is dangerous to apply our knowledge
of the dynamical structures in the solar system's dust cloud 
\cite{djxg94,lz99} directly to extrasolar systems
without first having considered the dust dynamics.
However, the example of the solar system also demonstrates that, once
we are able to detect more tenuous debris disks, perturbations from Neptune-mass
planets will be readily detectable, and it will even be possible to
detect structures associated with planets as small as the Earth.
In much the same way as it is not yet possible to detect the putative
planets around stars like Vega, the dust structures associated with
terrestrial planets may also be easier to detect than the planets themselves.

\section{Conclusions}
\label{s:conc}
This chapter has considered the types of structures seen in the dusty
debris disks of nearby stars (\S \ref{s:obs}) and how those structures can be used
to determine the layout of their planetary systems, in
terms of the distributions of both planetesimals and planets.
The text has dwelled on the successes of the models at explaining the observed
structures, because this illustrates the elements that are
essential to any debris disk model if the observations are to be successfully
explained (\S \ref{s:mod}), and because we are confident that we understand how a planet
would perturb a planetesimal belt in an idealised system comprised of just one planet
(\S \ref{s:plpl}) and to some extent how to extrapolate that to consider how the
planet would affect the observed dust disk (\S \ref{s:pldust}).
To summarise what we have learned:
\textbf{(i)} the axisymmetric structure of debris disks can mostly be explained by a
model in which dust is created in collisions in a narrow planetesimal belt and is
subsequently acted on by radiation forces;
\textbf{(ii)} the asymmetric structure of debris disks can mostly be explained by
secular and resonant gravitational perturbations from unseen planets acting on the
planetesimal belt and dust derived from it.

\begin{figure}
\centering
\begin{tabular}{c}
  \includegraphics[height=8.3cm]{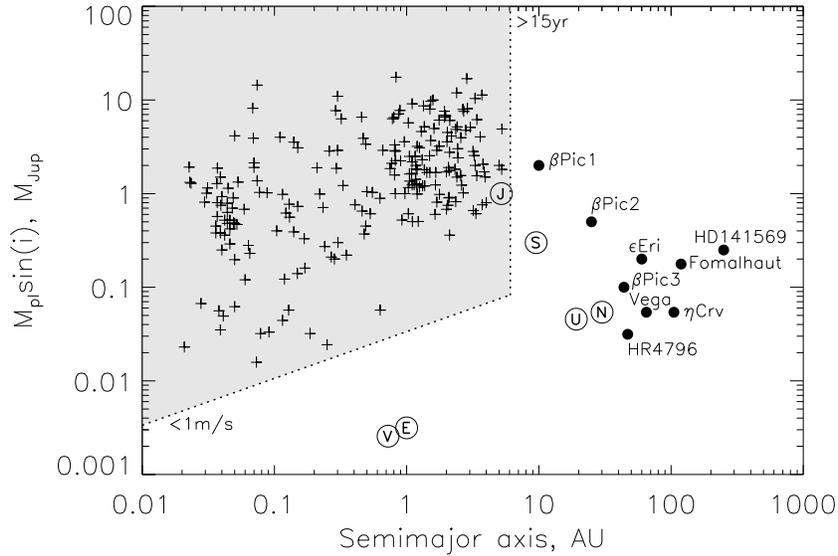}
\end{tabular}
\caption{Distribution of planet masses and semimajor axes.
Solar system planets are plotted as open circles, and those known from
radial velocity and transit studies with a plus (taken from the list on
http://exoplanets.eu dated 24 May 2007).
The shaded region shows the current limits of radial velocity surveys for
sun-like stars.
Debris disk planets inferred from disk structure (all awaiting
confirmation) are shown with filled circles.
References for the plotted planet parameters are:
HR4796 \cite{wdtf99},
$\epsilon$ Eridani \cite{ogmt00},
Vega \cite{wyat03},
HD141569 \cite{wyat05b},
$\eta$ Corvi \cite{wgdc05},
Fomalhaut \cite{quil06},
$\beta$ Pictoris \cite{fkl07},
although it should be noted that these parameters, particularly planet mass,
are often poorly constrained.}
\label{fig:exopl}
\end{figure}

Knowing the radial location of the planetesimal belts is important because this
demonstrates where in a protoplanetary disk grain growth must have
continued to km-sized planetesimals \cite{wd02}, and by analogy with
the solar system there is reason to believe that the location of the planetesimal
belts tells us indirectly the whereabouts of unseen planets, although it is worth
bearing in mind that there may be alternative explanations for
gaps in the planetesimal distribution related to the physics of the
protoplanetary disk.
Nevertheless, it appears that where we have the capability to look for detailed disk
structure, there is good correspondence between the asymmetric structures observed with
those expected if there are planets in these systems.
The modelling is also sufficiently advanced that the disk structure can be used
to infer information on the properties of the perturbing planets (such as the planet's
mass, orbit and even evolutionary history).
The planet properties which have been inferred in this way are particularly
exciting when compared with those of exoplanets discovered using the radial
velocity and transit techniques.
Figure \ref{fig:exopl} shows how the debris disk planets are similar to Uranus and
Neptune in the solar system, occupying a unique region of parameter space. 
This is possible because the large size of debris disks means that the planets perturbing them
are most often at large orbital radii, and it is easy for planets as small as Neptune
to impose structure on a debris disk.
There is also the tantalising possibility that in the future debris disk
structures can be used to identify planets analogous to the Earth and Venus
in extrasolar systems.

However, while it is incontrovertible that if there are planets present they
would impose structure on a disk, the question of whether we
have already seen these structures in extrasolar systems is still
a matter for debate.
In many cases the presence of an unseen planet
is the only explanation for the observed structures,
but that does not mean that it has to be the right explanation.
The problem is that it is hard to confirm that the planets are there, since
they lie beyond the reach of radial velocity studies (see Fig.~\ref{fig:exopl}).
Direct imaging could detect planets at this distance if they were a few times Jupiter
mass \cite{mhw03}, but not if they are Neptune mass.
Thus the onus is on the models to make other testable predictions,
and some of these have already been made (such as the orbital motion
of the clumpy structures, and the disk structures expected to be seen at
different wavelengths) and will be tested in the coming years.
If these planets are confirmed, their addition onto plots like that shown
in Figure \ref{fig:exopl} will be invaluable for constraining planet
formation models \cite{il04}.

It is also important to remember that this theory cannot yet predict
the quantities of small grains we would expect to see in any given disk.
There are too many uncertainties regarding the dust production mechanisms,
and it is possible that these processes may differ among stars with,
e.g., different dust compositions.
Applying dynamical models of the kind presented in \S \ref{s:mod} to a
greater number of resolved disk observations will help to understand these 
differences.
However, there is still the possibility that the problem is more
fundamental in a way which is best illustrated by the archetypal debris
disk Vega.
The observed mass loss rate from $\beta$ meteoroids in this
system is $2M_\oplus$/Myr, which indicates that this must be a transient,
rather than a steady state, component \cite{su05}.
It is thus possible that the small grain population in debris disks is
inherently stochastic, perhaps influenced by input from recent massive
collisions \cite{tele05}.
Fortunately it appears that the large grain component of the
majority of debris disks is evolving in steady state \cite{wssr07}
and so can be understood within the framework described in
this chapter, and the same is likely also true for the small grain component
(it is just the relative quantities of the different components
that is less certain).

However, the possibility must be considered that in some systems the
observed dust is transient in such a way that its origin will require
a significant overhaul to the models presented here.
For example, there are a few cases of sun-like stars surrounded by
hot dust (e.g., \S \ref{ss:axi}) which cannot be maintained
by steady state production in a massive asteroid given the age of the
stars \cite{wsgb07}.
It is not clear what the origin of the transient event producing the 
dust is.
However, it is known that the quantity of planetesimals in the inner solar
system has had a stochastic component, notably involving a large influx
$\sim 700$ Myr after the solar system formed in an event known as the
late heavy bombardment, the origin of which is thought to have been a
dynamical instability in the architecture of the giant planets
\cite{gltm05}.
So perhaps these systems are telling us about the more complex dynamics
of their planetary systems.
Given the complexity of planetary systems it seems inevitable that the models
presented in this chapter are just the start of a very exciting exploration of
the dynamics of extrasolar planetary systems.



\printindex

\begin{thebibliography}{99.}

\bibitem{absi06}
  O. Absil, et al:
  Astron. Astrophys. \textbf{452}, 237 (2006)
\bibitem{ardi04}
  D.R. Ardila, et al:
  Astrophys. J. \textbf{617}, L147 (2004)
\bibitem{ab05}
  J.-C. Augereau, H. Beust:
  Astron. Astrophys. \textbf{455}, 987 (2005)
\bibitem{anlp01}
  J.-C. Augereau, R.P. Nelson, A.M. Lagrange, J.C.B. Papaloizou, D. Mouillet:
  Astron. Astrophys. \textbf{370}, 447 (2001)
\bibitem{auma84}
  H.H. Aumann, et al: 
  Astrophys. J. \textbf{278}, L23 (1984)
\bibitem{bshb97}
  D. Barrado y Navascues, J.R. Stauffer, L. Hartmann, S.C. Balachandran: 
  Astrophys. J. \textbf{475}, 313 (1997)
\bibitem{bp93}
  D.E. Backman, F. Paresce: Debris disks.
  In: \textit{Protostars and Planets III},
  ed by E.H. Levy, J.I.Lunine
  (Tucson, Univ. Arizona Press 1993)
  pp 1253--1302
\bibitem{bm00}
  H. Beust, A. Morbidelli:
  Icarus \textbf{143}, 170 (2000)
\bibitem{bls79}
  J.A. Burns, P.L. Lamy, S. Soter:
  Icarus \textbf{40}, 1 (1979)
\bibitem{bryd06}
  G. Bryden, et al:
  Astrophys. J. \textbf{636}, 1098 (2006)
\bibitem{cpdb05}
  S. Charnoz, C. C. Porco, E. D\'{e}au, A. Brahic, J.N. Spitale,
  G. Bacques, K. Baillie:
  Science \textbf{310}, 1300 (2005)
\bibitem{clam03}
  M. Clampin, et al:
  Astron. J. \textbf{126}, 385 (2003)
\bibitem{dm05}
  A.T. Deller, S.T. Maddison:
  Astrophys. J. \textbf{625}, 398 (2005)
\bibitem{djxg94}
  S.F. Dermott, S. Jayaraman, Y.L. Xu, B.A.S. Gustafson, J.C. Liou:
  Nature, \textbf{369}, 719 (1994)
\bibitem{dd05}
  C.P. Dullemond, C. Dominik:
  Astron. Astrophys. \textbf{434}, 971 (2005)
\bibitem{fq07}
  P. Faber, A.C. Quillen:
  Mon. Not. Royal Astron. Soc. \textbf{submitted}, (2006)
\bibitem{ftpk00}
  R.S. Fisher, C.M. Telesco, R.K. Pi\~{n}a, R.F. Knacke, M.C. Wyatt:
  Astrophys. J. \textbf{532}, L141 (2000)
\bibitem{fv05}
  D.A. Fischer, J.A. Valenti:
  Astrophys. J. \textbf{622}, 1102 (2005)
\bibitem{fkl07}
  F. Freistetter, A.V. Krivov, T.L\"{o}hne:
  Astron. Astrophys. \textbf{466}, 389 (2007)
\bibitem{goli06}
  D.A. Golimowski, et al:
  Astron. J. \textbf{131}, 3109 (2006)
\bibitem{goli07}
  D.A. Golimowski, et al:
  In: \textit{Spirit of Lyot 2007},
  http://www.lyot2007.org (2007)
\bibitem{gltm05}
  R. Gomes, H.F. Levison, K. Tsiganis, A. Morbidelli:
  Nature \textbf{435}, 466 (2005)
\bibitem{gkm07}
  J.R. Graham, P.G. Kalas, B.C. Matthews:
  Astrophys. J. \textbf{654}, 595 (2007)
\bibitem{ghmj98}
  J.S. Greaves, et al.:
  Astrophys. J. \textbf{506}, L133 (1998)
\bibitem{gwhd04}
  J.S. Greaves, M.C. Wyatt, W.S. Holland, W.R.F. Dent:
  Mon. Not. Roy. Astron. Soc. \textbf{351}, L54 (2004)
\bibitem{ghwd05}
  J.S. Greaves, et al.:
  Astrophys. J. \textbf{619}, L187 (2005)
\bibitem{hm99}
  J.M. Hahn, R. Malhotra:
  Astron. J. \textbf{117}, 3041 (1999)
\bibitem{hllc00}
  S.R. Heap, D.J. Lindler, T.M. Lanz, R.H. Cornett,
  I. Hubeny, S.P. Maran, B. Woodgate:
  Astrophys. J. \textbf{539}, 435 (2000)
\bibitem{hgzw98}
  W.S. Holland, et al.:
  Nature \textbf{392}, 788 (1998)
\bibitem{hgdw03}
  W.S. Holland, et al.:
  Astrophys. J. \textbf{582}, 1141 (2003)
\bibitem{il04}
  S. Ida, D.N.C. Lin:
  Astrophys. J. \textbf{604}, 388 (2004)
\bibitem{jura04}
  M. Jura:
  Astrophys. J. \textbf{603}, 729 (2004)
\bibitem{kala05}
  P. Kalas:
  Astrophys. J. \textbf{635}, L169 (2005)
\bibitem{kj95}
  P. Kalas, D. Jewitt:
  Astron. J. \textbf{110}, 794 (1995)
\bibitem{kgc05}
  P. Kalas, J.R. Graham, M. Clampin:
  Nature \textbf{435}, 1067 (2005)
\bibitem{kgcf06}
  P. Kalas, J.R. Graham, M.C. Clampin, M.P. Fitzgerald:
  Astrophys. J. \textbf{637}, L57 (2006)
\bibitem{kfg07}
  P. Kalas, M.P. Fitzgerald, J.R. Graham:
  Astrophys. J. \textbf{661}, L85 (2007)
\bibitem{kso01}
  D.W. Koerner, A.I. Sargent, N.A. Ostroff:
  Astrophys. J. \textbf{560}, L181 (2001)
\bibitem{kris05}
  J.E. Krist, D.R. Ardila, D.A. Golimowski, M. Clampin, H.C. Ford:
  Astron. J. \textbf{129}, 1008 (2005)
\bibitem{kqls07}
  A.V. Krivov, M. Queck, T.L\"{o}hne, M. Sremcevic:
  Astron. Astrophys. \textbf{462}, 199 (2007)
\bibitem{kh03}
  M.J. Kuchner, M.J. Holman:
  Astrophys. J. \textbf{588}, 1100 (2003)
\bibitem{krb00}
  M.J. Kuchner, W. T. Reach, M. E. Brown:
  Icarus \textbf{145}, 44 (2000)
\bibitem{lp94}
  P.O. Lagage, E. Pantin:
  Nature \textbf{369}, 629 (1994)
\bibitem{lm07}
  C. Leinert, B. Moster:
  Astron. Astrophys. \textbf{472}, 335 (2007)
\bibitem{ll03}
  A. Li, J.I. Lunine:
  Astrophys. J. \textbf{590}, 368 (2003)
\bibitem{lolh06}
  S.-Y. Lin, N. Ohashi, J. Lim, P.T.P. Ho,
  M. Fukugawa, M. Tamura:
  Astrophys. J. \textbf{645}, 1297 (2006)
\bibitem{lz99}
  J.C. Liou, H.A. Zook:
  Astron. J. \textbf{118}, 580 (1999)
\bibitem{liu04}
  M.C. Liu:
  Science \textbf{305}, 1442 (2004)
\bibitem{malh95}
  R. Malhotra:
  Astron. J. \textbf{110}, 420 (1995)
\bibitem{mb98}
  V. Mannings, M.J. Barlow:
  Astrophys. J. \textbf{497}, 330 (1998)
\bibitem{mdvg06}
  K.A. Marsh, C.D. Dowell, T. Velusamy, K. Grogan, C.A. Beichman:
  Astrophys. J. \textbf{646}, L77 (2006)
\bibitem{mlpw07}
  R.G. Martin, S.H. Lubow, J.E. Pringle, M.C. Wyatt:
  Mon. Not. Royal Astron. Soc. \textbf{in press}, (2007)
\bibitem{mkw07}
  B.C. Matthews, P.G. Kalas, M.C. Wyatt:
  Astrophys. J. \textbf{in press}, (2007)
\bibitem{mhw03}
  S.A. Metchev, L.A. Hillenbrand, R.J. White:
  Astrophys. J. \textbf{582}, 1102 (2003)
\bibitem{mm02}
  A. Moro-Mart\'{i}n, R. Malhotra:
  Astron. J. \textbf{124}, 2305 (2002)
\bibitem{mm05}
  A. Moro-Mart\'{i}n, R. Malhotra:
  Astrophys. J. \textbf{633}, 1150 (2005)
\bibitem{mwm05}
  A. Moro-Mart\'{i}n, S. Wolf, R. Malhotra:
  Astrophys. J. \textbf{621}, 1079 (2005)
\bibitem{mc05}
  R.A. Murray-Clay, E.I. Chiang:
  Astrophys. J. \textbf{619}, 623 (2005)
\bibitem{md99}
  C.D. Murray, S.F. Dermott:
  \textit{Solar System Dynamics}
  (Cambridge University Press, Cambridge 1999)
\bibitem{ogmt00}
  L.M. Ozernoy, N.N. Gorkavyi, J.C. Mather, T.A. Taidakova:
  Astrophys. J. \textbf{537}, L147 (2000)
\bibitem{pjl05}
  P. Plavchan, M. Jura, S.J. Lipscy:
  Astrophys. J. \textbf{631}, 1161 (2005)
\bibitem{pgc06}
  C.J. Poulton, J.S. Greaves, A.C. Cameron:
  Mon. Not. Royal Astron. Soc. \textbf{372}, 53 (2006)
\bibitem{qt02}
  A.C. Quillen, S. Thorndike:
  Astrophys. J. \textbf{578}, L149 (2002)
\bibitem{quil06}
  A.C. Quillen:
  Mon. Not. Royal Astron. Soc. \textbf{372}, L14 (2006)
\bibitem{rwm05}
  A. Roberge, A.J. Weinberger, E.M. Malumuth:
  Astrophys. J. \textbf{622}, 1151 (2005)
\bibitem{rsss94}
  F. Roques, H. Scholl, B. Sicardy, B.A. Smith:
  Icarus \textbf{108}, 37 (1994)
\bibitem{ssbk99}
  G. Schneider, et al:
  Astrophys. J. \textbf{513}, L127 (1999)
\bibitem{ssh05}
  G. Schneider, M.D. Silverstone, D.C. Hines:
  Astrophys. J. \textbf{629}, L117 (2005)
\bibitem{schn06}
  G. Schneider, et al:
  Astrophys. J. \textbf{650}, 414 (2006)
\bibitem{st84}
  B.A. Smith, R.J. Terrile:
  Science \textbf{226}, 1421 (1984)
\bibitem{stap04}
  K.R. Stapelfeldt, et al:
  Astrophys. J. Suppl. \textbf{154}, 458 (2004)
\bibitem{sc06}
  L.E. Strubbe, E.I. Chiang:
  Astrophys. J. \textbf{648}, 652 (2006)
\bibitem{su05}
  K.Y.L. Su, et al:
  Astrophys. J. \textbf{628}, 487 (2005)
\bibitem{su06}
  K.Y.L. Su, et al:
  Astrophys. J. \textbf{653}, 675 (2006)
\bibitem{tin96}
  H. Tanaka, S. Inaba, K. Nakazawa:
  Icarus \textbf{123}, 450 (1996)
\bibitem{tele00}
  C.M. Telesco, et al:
  Astrophys. J. \textbf{530}, 329 (2000)
\bibitem{tele05}
  C.M. Telesco, et al:
  Nature \textbf{433}, 133 (2005)
\bibitem{ta07}
  P. Th\'{e}bault, J.C. Augereau:
  Astron. Astrophys. \textbf{in press}, astro-ph$/$07060344 (2007)
\bibitem{tab03}
  P. Th\'{e}bault, J.C. Augereau, H. Beust:
  Astron. Astrophys. \textbf{408}, 775 (2003)
\bibitem{tb01}
  C.A. Trujillo, M.E. Brown:
  Astrophys. J. \textbf{554}, L95 (2001)
\bibitem{wrbz00}
  A.J. Weinberger, R.M. Rich, E.E. Becklin, B. Zuckerman,
  K. Matthews:
  Astrophys. J. \textbf{544}, 937 (2000)
\bibitem{wnlb04}
  J.P. Williams, J. Najita, M.C. Liu, S. Bottinelli, J.M. Carpenter,
  L.A. Hillenbrand, M.R. Meyer, D.R. Soderblom:
  Astrophys. J. \textbf{604}, 414 (2004)
\bibitem{whkh02}
  D.J. Wilner, M.J. Holman, M.J. Kuchner, P.T.P. Ho:
  Astrophys. J. \textbf{569}, L115 (2002)
\bibitem{wisd80}
  J. Wisdom:
  Astron. J. \textbf{85}, 1122 (1980)
\bibitem{wyat99}
  M.C. Wyatt:
  Signatures of Planets in Circumstellar Disks.
  Ph.D. Thesis, University of Florida, Gainesville (1999)
\bibitem{wyat03}
  M.C. Wyatt:
  Astrophys. J. \textbf{598}, 1321 (2003)
\bibitem{wyat05a}
  M.C. Wyatt:
  Astron. Astrophys. \textbf{433}, 1007 (2005)
\bibitem{wyat05b}
  M.C. Wyatt:
  Astron. Astrophys. \textbf{440}, 937 (2005)
\bibitem{wyat06}
  M.C. Wyatt:
  Astrophys. J. \textbf{639}, 1153 (2006)
\bibitem{wd02}
  M.C. Wyatt, W.R.F. Dent: 
  Mon. Not. Roy. Astron. Soc. \textbf{334}, 589 (2002)
\bibitem{wdtf99}
  M.C. Wyatt, S.F. Dermott, C.M. Telesco, R.S. Fisher,
  K. Grogan, E.K. Holmes, R.K. Pi\~{n}a:
  Astrophys. J. \textbf{527}, 918 (1999)
\bibitem{whgd03}
  M.C. Wyatt, W.S. Holland, J.S. Greaves, W.R.F. Dent:
  Earth Moon and Planets \textbf{92}, 423 (2003)
\bibitem{wgdc05}
  M.C. Wyatt, J.S. Greaves, W.R.F. Dent, I.M. Coulson:
  Astrophys. J. \textbf{620}, 492 (2005)
\bibitem{wsgb07}
  M.C. Wyatt, R. Smith, J.S. Greaves, C.A. Beichman, G. 
  Bryden, C.M. Lisse:
  Astrophys. J. \textbf{658}, 569 (2007)
\bibitem{wssr07}
  M.C. Wyatt, R. Smith, K.Y.L. Su, G.H. Rieke, 
  J.S. Greaves, C.A. Beichman, G. Bryden:
  Astrophys. J. \textbf{663}, 365 (2007)

\end{thebibliography}
\end{document}